\documentclass[final,3p]{elsarticle}
\usepackage{graphicx}
\usepackage{epsfig}
\usepackage{epstopdf}
\usepackage{subfig}

\usepackage{amssymb}
\usepackage{amsthm}
\usepackage{setspace}
\usepackage{lineno}
\usepackage{appendix}
\usepackage{hyperref}
\usepackage[all]{hypcap}
\journal{elsevier}

\begin{document}
\onecolumn
%\tableofcontents{}
\setcounter{tocdepth}{2}
%\addcontentsline{toc}{chapter}{List of Figures}
%\listoffigures

\newpage

\begin{frontmatter}
%\linenumbers



\title{A 15 GSa/s, 1.5 GHz Bandwidth Waveform Digitizing ASIC}% in 0.13~$\mu$m~CMOS}

\author[efi]{Eric Oberla}\corref{cor1}\ead{ejo@uchicago.edu}
%\author[efi]{{Herv\'{e} Grabas}\fnref{Grabas_PAddress}}
\author[efi]{{Jean-Francois Genat}\fnref{Genat_PAddress}}
\author[efi]{{Herv\'{e} Grabas}\fnref{Grabas_PAddress}}
\author[efi]{Henry Frisch}
\author[hawaii]{{Kurtis Nishimura}\fnref{kurtis_pAddress}}
\author[hawaii]{Gary Varner}
\address[efi]{Enrico Fermi Institute, University of Chicago; 5640
  S. Ellis Ave., Chicago IL, 60637}
\address[hawaii]{University of Hawai'i at Manoa; Watanabe Hall, 2505
  Correa Rd., Honolulu HA}

\cortext[cor1]{Corresponding author}

\fntext[Genat_PAddress]{Present address, LPNHE, CNRS/IN2P3,
Universit\'{e}s Pierre et Marie Curie and Denis Diderot, T12 RC,
4 Place Jussieu 75252 Paris CEDEX 05, France}
\fntext[Grabas_PAddress]{Present address, CEA/IRFU/SEDI; CEN Saclay-Bat141
F-91191 Gif-sur-Yvette CEDEX, France}
%\fntext[Genat_PAddress]{Present address, LPNHE, CNRS/IN2P3,
%Universit\'{e}s Pierre et Marie Curie and Denis Diderot, T12 RC,
%4 Place Jussieu 75252 Paris CEDEX 05, France}
\fntext[kurtis_pAddress]{Present address, SLAC National Accelerator Laboratory,
2575 Sand Hill Road, Menlo Park, CA 94025}

\begin{abstract}

The PSEC4 custom integrated circuit was designed for the recording of
fast waveforms for use in large-area time-of-flight detector systems. 
The ASIC has been fabricated using the IBM-8RF
0.13 $\mu$m CMOS process.   On each of 6~analog channels, PSEC4 employs
a switched capacitor array (SCA) 256~samples deep, a 
ramp-compare ADC with 10.5~bits of DC dynamic range, and a
serial data readout with the capability of
region-of-interest windowing to reduce dead time.  
The sampling rate can be
adjusted between 4~and 15~Gigasamples/second~[GSa/s] on all channels
and is servo-controlled on-chip with a low-jitter delay-locked loop (DLL). The input signals are passively coupled
on-chip with a -3~dB analog bandwidth of 1.5 GHz.  The power consumption in quiescent sampling mode 
is less than 50~mW/chip; at a sustained trigger and readout rate of 50~kHz the chip
draws 100~mW.  
After fixed-pattern pedestal subtraction, the uncorrected integral non-linearity is 0.15\% 
over an 750~mV dynamic range. With a linearity correction, a full 1~V signal voltage range is available.
The sampling timebase has a fixed-pattern non-linearity with an RMS of 13$\%$, 
which can be corrected for precision waveform feature extraction and timing.

\end{abstract}


\end{frontmatter}

%\linenumbers
%\doublespacing
\section{Introduction}
\label{introduction}
We describe the design and performance of PSEC4, a $\geq$10~Gigasample/second~[GSa/s] waveform sampling
and digitizing Application Specific Integrated Circuit (ASIC) fabricated in the IBM-8RF 0.13~$\mu$m 
complementary metal-oxide-semiconductor (CMOS) technology. This compact `oscilloscope-on-a-chip' 
is designed for the recording of radio-frequency (RF) transient waveforms with signal bandwidths 
between 100~MHz and 1.5~GHz. 

\subsection{Background}
The detection of discrete photons and high-energy particles 
is the basis of a wide range of commercial and
scientific applications. 
In many of these applications, the relative arrival time of an incident photon or particle is best measured
by extracting features from the full waveform at the detector output~\cite{JF_NIM, Vavra}. 
Additional benefits of front-end waveform sampling include the detection of pile-up events and
the ability to filter noise or poorly formed pulses.

For recording `snapshots' of transient waveforms, switched capacitor array (SCA)
analog memories can be used to sample a limited time-window  at a relatively high rate, 
but with a latency-cost of a slower readout speed~\cite{kleinSCA, gunther}.
These devices are well suited for triggered-event applications, as in many high energy physics 
experiments, in which some dead-time on each channel is acceptable. With modern CMOS integrated
circuit design, these SCA sampling chips can be compact, low power, and have a relatively low cost per channel~\cite{gunther}.  

Over the last decade, %
sampling rates in SCA waveform sampling ASICs 
have been pushed to several GSa/s with analog bandwidths of 
several hundred MHz up to $\sim$1~GHz~\cite{ritt,blab}.
As a scalable front-end readout option coupled with the advantages of waveform sampling, 
these ASICs have been used in a wide range of experiments;
such as high-energy physics colliders~\cite{blab},  gamma-ray
astronomy~\cite{delagnes, target}, 
high-energy neutrino detection~\cite{kleinfelder_GHz,varner_radio}, and rare decay searches~\cite{DSC,DRS4}. 

\subsection{Motivation}
A natural extension to the existing waveform sampling ASICs is to push
design parameters that are inherently fabrication-technology limited.
Parameters such as sampling rate and analog bandwidth are of particular
interest considering the fast risetimes ($\tau$$_r$~$\sim$~60~-~500~ps)  and pulse widths 
(FWHM~$\sim$ 200~ps~-~1~ns) of commercially available and novel technologies of micro-channel plate (MCP) 
and silicon photomultipliers~\cite{fastMCP, elagin,  sipm}. These and other fast
photo-optical or RF devices require electronics matched to the speed of the signals.

The timing resolution of discrete waveform sampling is intuitively dependent on three primary 
factors as described by Ritt\footnote{Assuming Shannon-Nyquist is fulfilled}~\cite{rittsnr}:
\begin{equation}
\sigma_t \: \propto \: \frac{\tau_r}{(SNR) \sqrt{N_{samples}}}
\end{equation}
where \textit{SNR} is the signal-to-noise ratio of the pulse, 
$\tau$$_r$ is the 10-90\% rise-time of the pulse, and
\textit{N$_{samples}$} is
the number of independent samples on the rising edge within time $\tau$$_r$.
The motivation for oversampling above the Nyquist limit is that errors due to uncorrelated noise, 
caused both by random time jitter
and charge fluctuations, are reduced by increasing the rising-edge sample size.
Accordingly, in order to preserve
the timing properties of analog signals from a fast detector, the waveform recording electronics 
should 1) be low-noise, 2) match the signal bandwidth, 
and 3) have a fast sampling rate relative to the signal rise-time.

\subsection{Towards 0.13 $\mu$m CMOS}

The well-known advantages of reduced transistor feature size 
include higher clock speeds, greater circuit
density, lower parasitic capacitances, and lower power dissipation per circuit~\cite{dennard}. 
The sampling rate and analog bandwidth of waveform sampling ASICs, which depend on clock speeds,
parasitic capacitances, and interconnect lengths, 
are directly  enhanced by moving to a smaller CMOS technology.
Designing in a smaller technology also allows 
clocking of an on-chip analog-to-digital converter (ADC) at a faster rate, reducing the chip dead-time.

With the advantages of reduced transistor feature sizes also comes increasingly challenging
analog design issues. One issue is the increase of leakage current. Leakage is enhanced by
decreased source-drain channel lengths, causing subthreshold leakage (V$_{GS}$ \textless~V$_{TH}$),
and decreased gate-oxide thickness, which promotes gate-oxide tunneling~\cite{taur}. Effects of leakage
include increased quiescent power dissipation and potential non-linear effects when storing analog voltages.

Another design issue of deeper sub-micron technologies is the
reduced dynamic range~\cite{taur}. The available voltage range is given by (V$_{DD}$-V$_{TH}$), where V$_{DD}$ is the
supply voltage and V$_{TH}$ is the threshold, or `turn-on', voltage for a given transistor. For technologies
above 0.1 $\mu$m, the (V$_{DD}$-V$_{TH}$) 
range is decreased with downscaled feature sizes to reduce high-field effects in the gate-oxide~\cite{taur}. 
In the 0.13~$\mu$m  CMOS process, 
the supply voltage V$_{DD}$ is 1.2~V 
and the values of V$_{TH}$ range from 0.42~V for a minimum-size transistor (gate length 120 nm) to roughly 0.2~V 
for a large transistor (5~$\mu$m)~\cite{ibm, mosis}.

%\begin{equation}
%f_T \: \propto \: \frac{\mu}{L^2} (V_{GS}-V_T)
%\end{equation}

\begin{figure}[]
%\hspace{-40 pt}
\centering
\includegraphics[scale=.35]{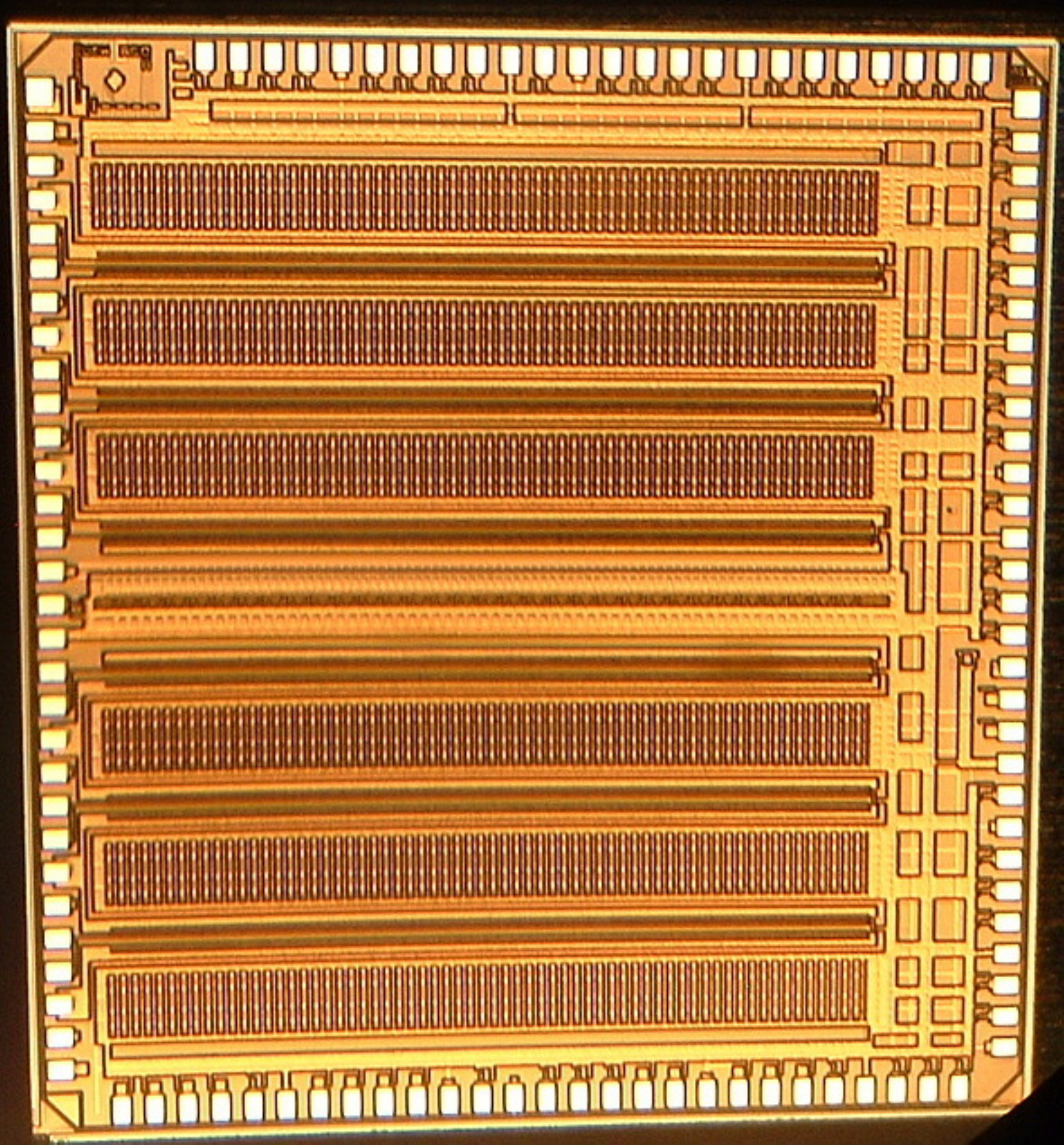}
\vspace{ -5 pt}
\caption[die photo]{The fabricated PSEC4 chip. The die dimensions are 4$\times$4.2 mm$^2$. }
\label{die}
\end{figure}

The potential of waveform sampling design in 0.13~$\mu$m CMOS was shown 
with two previous ASICs. A waveform sampling prototype, PSEC3, achieved a sampling rate of 15~GSa/s 
and showed the possibility of analog bandwidths above 1~GHz~\cite{psec3}. 
Leakage and dynamic range studies were also performed with this chip.
In a separate 0.13 $\mu$m ASIC, fabricated as a test-structure chip called CHAMP, 
a 25~GSa/s sampling rate
rate was achieved using low~V$_{TH}$ transistors~\cite{champ}. 
The performance and limitations of these chips led 
to the optimized design of the PSEC4 waveform digitizing ASIC.
The fabricated PSEC4 die is shown in Figure~\ref{die}.

In this paper, we describe the PSEC4 architecture ($\S$\ref{architecture}), 
and experimental performance ($\S$\ref{performance}).
%and a first application 
%to the front-end readout of large-area, picosecond resolution photodetectors ($\S$\ref{application}).

%%%%%%%%%%%%%%%%%%%%%%%%%%%%%%%%%%%%%%%%%%%%%%%%%
\section{Architecture}
\label{architecture}

\begin{figure*}[]
\vspace{-40 pt}
%\hspace{-0 pt}
\centering
\includegraphics[scale=1.4]{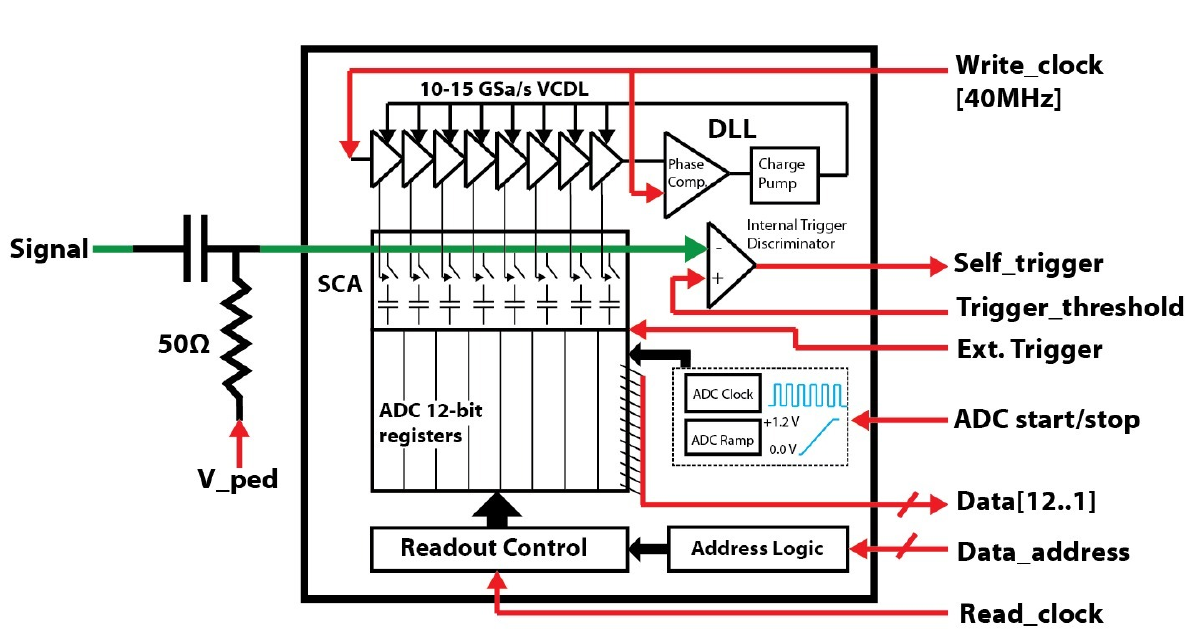}
\vspace{-5 pt}
\caption[Overview block diagram of PSEC-4 control]{A block diagram of
PSEC4 functionality. The RF-input signal is AC coupled and terminated in 50$\Omega$ off-chip. The digital signals 
(listed on right) are interfaced with an FPGA for PSEC4 control. 
A 40~MHz write clock is fed to the chip and up-converted to $\sim$10~GSa/s with a 
256-stage voltage-controlled delay line (VCDL). (For clarity, only 8 of the 256 cells 
and 1 of the 6 channels are illustrated). A `write strobe' signal is sent from each 
stage of the VCDL to the corresponding sampling cell in each channel. The
write strobe passes the VCDL-generated sampling rate to the sample-and-hold switches of each SCA cell.
Each cell is made from a switched capacitor sampling cell and integrated ADC counter, as shown in Figure~\ref{psec_cell}.
The trigger signal ultimately comes from the FPGA, in which sampling on every channel 
is halted and all analog samples are digitized. The on-chip ramp-compare ADC is run with a global
analog ramp generator and 1~GHz clock that are distributed to each cell. 
Once digitized, the addressed data are serially sent off-chip on a 12-bit bus clocked at up to 80~MHz.}
\label{psec4block}
\end{figure*}

\begin{figure*}[]
%\vspace{-300 pt}
\centering
%\hspace{20 pt}
\includegraphics[scale=.63]{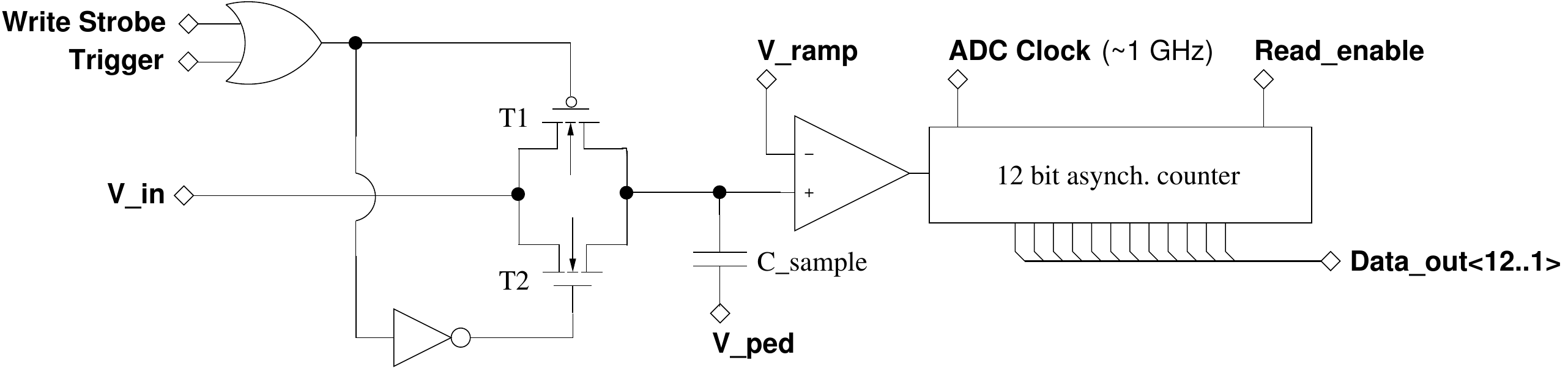}
\caption[psec cell]{Simplified schematic of the `vertically integrated' PSEC4 cell structure. 
The sampling cell input, V$_{in}$,
is tied to the on-chip 50$\Omega$ input microstrip line. Transistors T1 and T2 form a dual-CMOS write switch that facilitates
the sample-and-hold of V$_{in}$ on C$_{sample}$, a 20~fF capacitance referenced to \textit{V\_{ped}}. The switch is toggled by the VCDL write strobe while sampling 
(Figure~\ref{psec4block}) or an ASIC-global trigger signal when an event is to be digitized. 
When the ADC is initiated, a global 0.0-1.2~V
analog voltage ramp is sent to all comparators, which digitizes the voltage on 
C$_{sample}$ using a fast ADC clock and 12-bit counter. 
To send the digital data off-chip, the register is addressed using \textit{Read\_enable}.}
\label{psec_cell}
\end{figure*}

An overview of the PSEC4 architecture and functionality is shown in Figure~\ref{psec4block}. 
 A PSEC4 channel is a linear array
of 256 sample points and a threshold-level trigger discriminator. Each sample point in the array is made from a 
switched capacitor sampling cell 
and an integrated ADC circuit  
as shown in Figure~\ref{psec_cell}. 

To operate the chip, a field-programmable gate array (FPGA) is used to provide timing control, clock generation, 
readout addressing, data management, and general 
configurations to the ASIC. Several analog voltage controls are also required for operation, and are 
provided by commercial digital-to-analog converter (DAC) chips.

%The chip architecture is well suited for triggered event applications, as the digitization and readout induced latency. 

Further details of the chip architecture, including timing generation~($\S$\ref{timing generation}) 
sampling and triggering~($\S$\ref{sampling and triggering}), 
 and analog-to-digital conversion~($\S$\ref{analog to digital}), 
%and  data readout~($\S$\ref{readout}), 
are outlined in the following sections. 

\subsection{Timing Generation}
\label{timing generation}
\begin{figure}[t!]
%\hspace{-30 pt}
\centering
\includegraphics[scale=0.35]{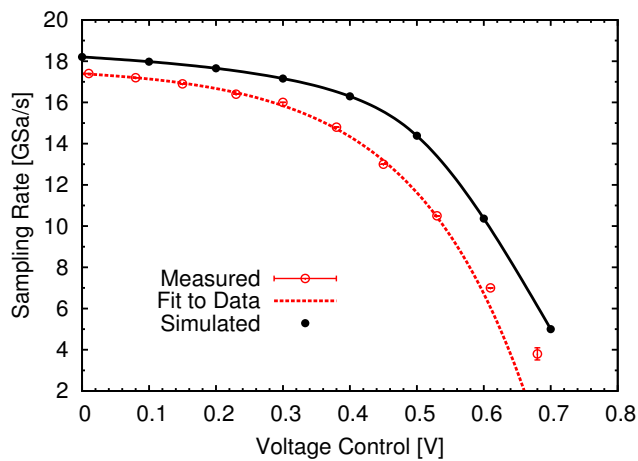}
\vspace{-20 pt}
\caption[Sampling rate]{Sampling rate as a function of VCDL voltage control. Good agreement is shown between post layout 
simulation and actual values. Rates up to 17.5~GSa/s are achieved with the free-running PSEC4 VCDL. 
%When operating the VCDL without feedback, the control voltage is explicitly 
%set and the sampling rate is given by $ 17.7 \: (1 - 0.018 \: \exp(5.91 \cdot V_{control}))$ [GSa/s].
%Typically, the servo-locking will be enabled and the VCDL is run as a delay-locked loop (DLL). In this case, the sampling rate
%is automatically set by the input write clock frequency.
}
\label{fig:samplerate}
\end{figure}

The sampling signals are generated with a 256-stage Voltage-Controlled Delay Line (VCDL), 
in which the individual stage time delay is adjustable by two complementary voltage controls.  
There is a single VCDL that distributes the timing signals to the entire chip. Each stage
in the VCDL is an RC delay element made from a CMOS current-starved inverter.
The inverse of the time delay between stages 
sets the sampling rate. Rates of up to 17.5~GSa/s are possible with PSEC4 as shown in Figure~\ref{fig:samplerate}. 
When operating the VCDL without feedback, the control voltage is explicitly 
set and the sampling rate is approximately given by %$ 17.7 \: (1 - 0.018 \: \exp(5.91 \cdot V_{control}))$
\begin{equation}
% 18 \: (1 -  1.8\times10^{-2} \: e^{6 \cdot V_{control}}) \: \: [GSa/s].
 18 -  0.3 \: e^{6 \cdot V_{control}} \: \: [GSa/s].
\end{equation}
Typically, the servo-locking will be enabled and the VCDL is run as a delay-locked loop (DLL). In this case, the sampling rate
is automatically set by the input write clock frequency.
The stability of the sampling rate is negatively correlated with the slope magnitude as the VCDL becomes increasingly 
sensitive to noise. The slowest stable sampling rate is $\sim$4~GSa/s.

A `write strobe' signal is sent from each stage of the VCDL to the corresponding sampling cell in each channel. The
write strobe passes the VCDL-generated sampling rate to the sample-and-hold switch of the cell as shown in 
Figure~\ref{psec_cell}.
To allow the sample cell enough time to fully charge or discharge when sampling, the write strobe is extended to a 
fixed duration of 8$\times$ the individual VCDL delay stage. In sampling mode, 
a `sampling block' made of 8 adjacent SCA sampling cells continuously tracks the input signal.

To servo-control the VCDL at a specified sampling rate and to
compensate for temperature effects and power supply variations, 
the VCDL can be delay-locked on chip. 
The VCDL forms a delay-locked loop (DLL) when this servo-controlled feedback
is enabled.
The servo-control circuit is made of a dual phase comparator and charge pump circuit to lock both the rising and falling edges of 
the write clock at a fixed one-cycle latency~\cite{dll}. 
 A loop-filter capacitor is installed externally to tune the DLL stability. 

With this DLL architecture, a write clock with frequency f$_{in}$ is provided to the chip, and 
the sampling is started automatically after
a locking time of several seconds. The nominal sampling rate in GSa/s is set by 0.256$\cdot$f$_{in}$~[MHz], and
the sampling buffer depth in nanoseconds is given by 10$^3$/f$_{in}$~[MHz$^{-1}$]. A limitation of the PSEC4 design
is the relatively small recording depth at high sampling rates due to the buffer size of 256~samples.

\subsection{Sampling and Triggering}
\label{sampling and triggering}
A single-ended, 256-cell SCA was designed and implemented on each channel of PSEC4. 
Each sampling cell circuit is made from a dual CMOS write switch and a metal-insulator-metal
sampling capacitor as shown in Figure~\ref{psec_cell}. With layout parasitics, this capacitance is
effectively 20~fF.
During sampling, the write switch is toggled by the write 
strobe from the VCDL. To record an event, an external trigger, typically from an FPGA, overrides 
the sampling and opens all write switches, 
holding the analog voltages on the capacitor for the ADC duration ($\leq$4~$\mu$s).  
Triggering interrupts the sampling on every channel, and is held until the selected data are digitized and read out.
%When triggered, the sampling is asynchronously halted. This corrupts the voltage on the 7 sample cells 
%that were in the process of sampling, reducing the effective number of PSEC4 samples to 249.  

The PSEC4 has the capability to output a threshold-level trigger bit on each channel. 
The internal trigger is made from a fast comparator, 
which is referenced to an external threshold level, and
digital logic to latch and reset the trigger circuit. 
To form a PSEC4 trigger, the self-trigger bits are sent to the FPGA, which returns a global trigger signal back to the chip. 
In the internal trigger mode, the trigger round-trip time is 15-20~ns (depending on FPGA algorithm), which allows for the recording of a waveform before it is overwritten at 10 GSa/s.

\subsection{ADC}
\label{analog to digital}
Digital conversion of the sampled waveforms is done on-chip with a single ramp-compare ADC that is parallelized over the entire ASIC\footnote{An
overview of this ADC architecture can be found in reference~\cite{rampadc}.}. 
Each sample cell has a dedicated comparator and 12~bit counter as shown in Figure~\ref{psec_cell}. 
In this architecture, the comparison between each sampled voltage (\textit{V$_{sample}$}) and a global ramping voltage (\textit{V$_{ramp}$}), 
 controls the clock enable of a 12-bit counter. When V$_{ramp}$~\textgreater~V$_{sample}$, the counter clocking is disabled, and
the 12-bit word, which has been encoded by the ADC clock frequency and the ramp duration below \textit{V$_{sample}$}, 
is latched and ready for readout.  

Embedded in each channel is a 5-stage ring oscillator that generates a fast digital ADC clock,
adjustable between 200~MHz and 1.4~GHz.  
The ADC conversion time, power consumption, and resolution may be configured by adjusting the ramp slope or by tuning the 
ring oscillator frequency.

%%%%%%%%%%%%%%%%%%%%%%%%%%%%%%%%%%%%%%%%%%%%%
%%%%%%%%%%%%%%%%%%%%%%%%%%%%%%%%%%%%%%%%%%%%%
\section{Performance}
\label{performance}

\begin{figure}[t!]
%\hspace{-10 pt}
\centering
\includegraphics[scale=.5]{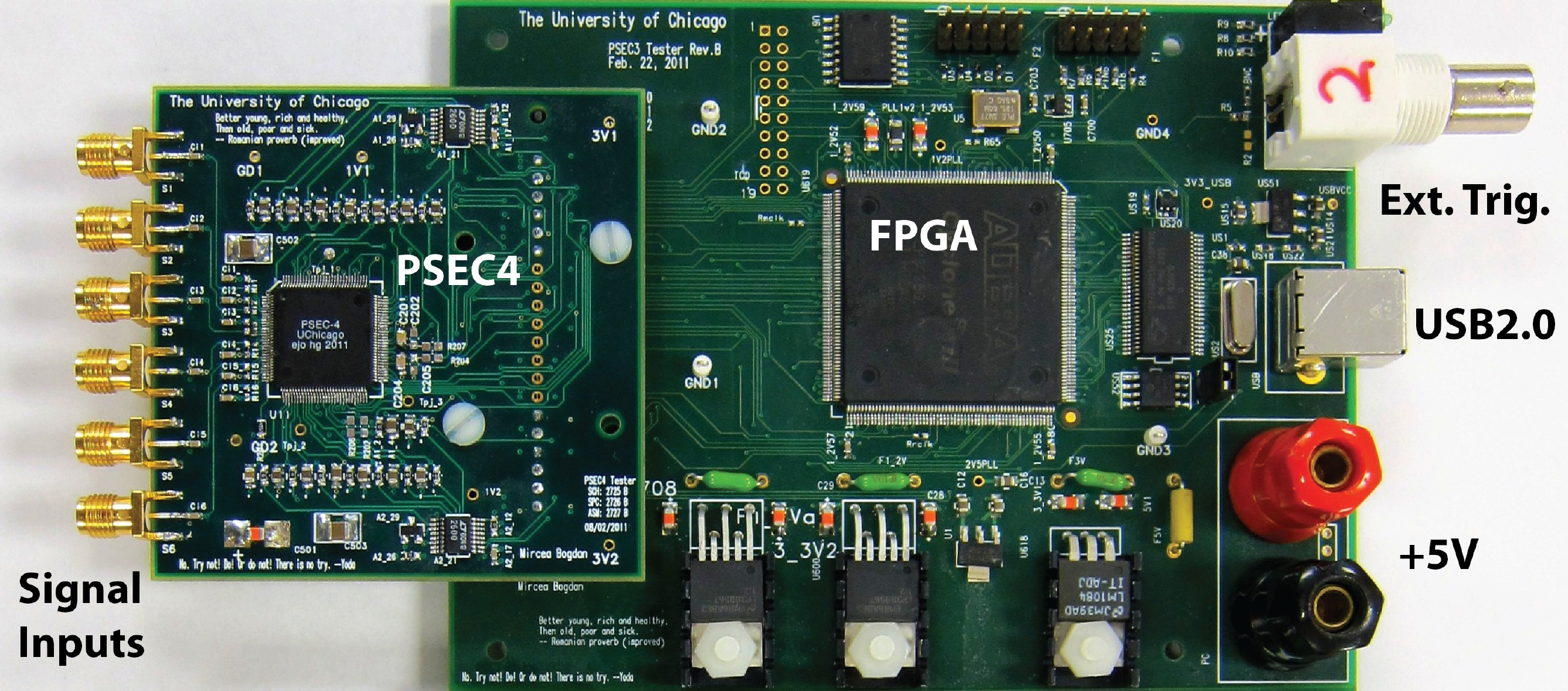}
\vspace{-5 pt}
\caption[psec4 eval]{The PSEC4 evaluation board. The board uses a Cyclone III Altera FPGA (EP3C25Q240) and a USB~2.0 PC interface. Custom firmware and acquisition software were developed for overall board control. The board uses +5~V power and draws $\textless$400~mA, either from a DC supply or the USB interface.}
\label{eval}
\end{figure}

Measurements of the PSEC4 performance have been made with several
chips on custom evaluation boards shown in Figure~\ref{eval}. The
sampling rate was fixed at a nominal rate of 10.24~GSa/s.
Here we report on bench measurements of
linearity~($\S$\ref{linearity}),
analog leakage~($\S$\ref{leakage_section}),
noise~($\S$\ref{noise}),
power~($\S$\ref{power}),
frequency response~($\S$\ref{freq}),
sampling calibrations~($\S$\ref{timebase}),
and waveform timing~($\S$\ref{timing}). 
A summary table of the PSEC4 performance is shown in $\S$\ref{perf_table}.
%The sampling rate was fixed at a nominal rate of 10.24 GSa/s
%for these measurements.

\subsection{Linearity and Dynamic Range}
\label{linearity}

The signal voltage range is limited by the 1.2~V core voltage of the 0.13~$\mu$m CMOS process~\cite{ibm}.
To enable the recording of signals with pedestal levels that exceed this range, the input is AC coupled and a DC offset
is added to the 50~$\Omega$ termination. This is shown in the  Figure~\ref{psec4block} block diagram, in which the DC
offset is designated by \textit{V\_ped}. 
The offset level is tuned to match the input signal voltage range to that of PSEC4.

\begin{figure}[t!]
\hspace{-46 pt}
\centering
\includegraphics[scale=.34]{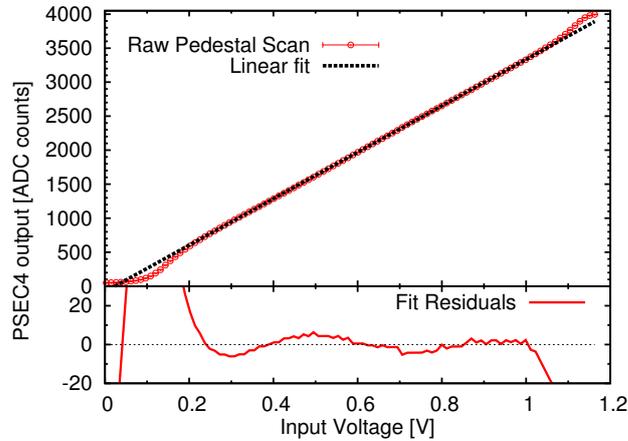}
\vspace{-25 pt}
\caption[ADC linearity]{DC response of the device running in 12 bit mode. The data are an average response of all 256 cells from a single channel.
The upper plot shows raw data (red points) and a 
linear fit over the the same dynamic range (dotted black line, slope of 4 counts/mV). The fit residuals are shown in the lower plot. }
%A differential non-linearity (DNL) of better than 0.15$\%$ is observed for input signals between 0.2~V and 1.0~V before any calibrations. }
\label{adclinear}
\end{figure}

The PSEC4-channel response to a linear pedestal scan is shown in Figure~\ref{adclinear}. 
This is the average DC response over all 256 cells in a channel. 
A signal voltage range of 1~V is shown, as input signals between 100~mV 
and 1.1~V are fully coded with 12~bits. 
An integral non-linearity (INL) of better than 0.15$\%$ is shown for most of that range.
The non-linearity and limited DC signal range near the voltage rails are 
due to transistor threshold issues in the comparator circuit.
 
The DNL of this response, shown by the linear fit residuals in Figure~\ref{adclinear}, can be corrected by
creating an ADC count-to-voltage look-up-table (LUT) that maps the input voltage to the PSEC4 output code. 
The raw PSEC4 data is converted to voltage and `linearized' with a channel-averaged LUT.

%%%%%%%%%%%%%%%

\begin{figure}[t!]
\centering
\includegraphics[scale=.44]{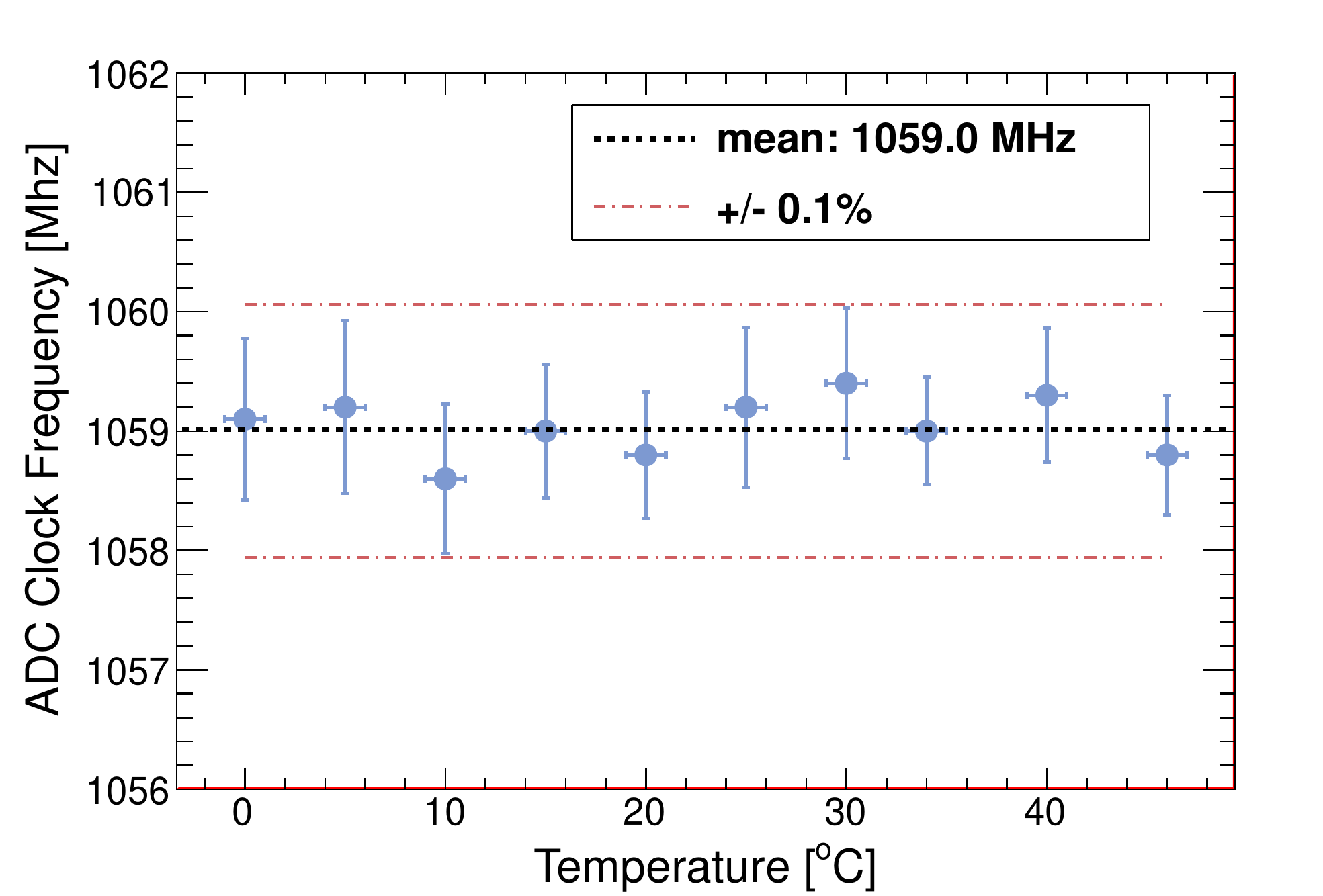}
\caption{Ring oscillator clock stability over temperature. This clock  is stabilized with a 
servo-control algorithm in the FPGA that adjusts the DAC oscillator voltage controls. 
With this feedback, the ring oscillator frequency is held within 0.1\% of the nominal 1.059 GHz over the tested
temperature range.}
\label{ro_temp}
\end{figure}

\begin{figure}[t!]
\centering
\subfloat[]{\includegraphics[scale=.44]{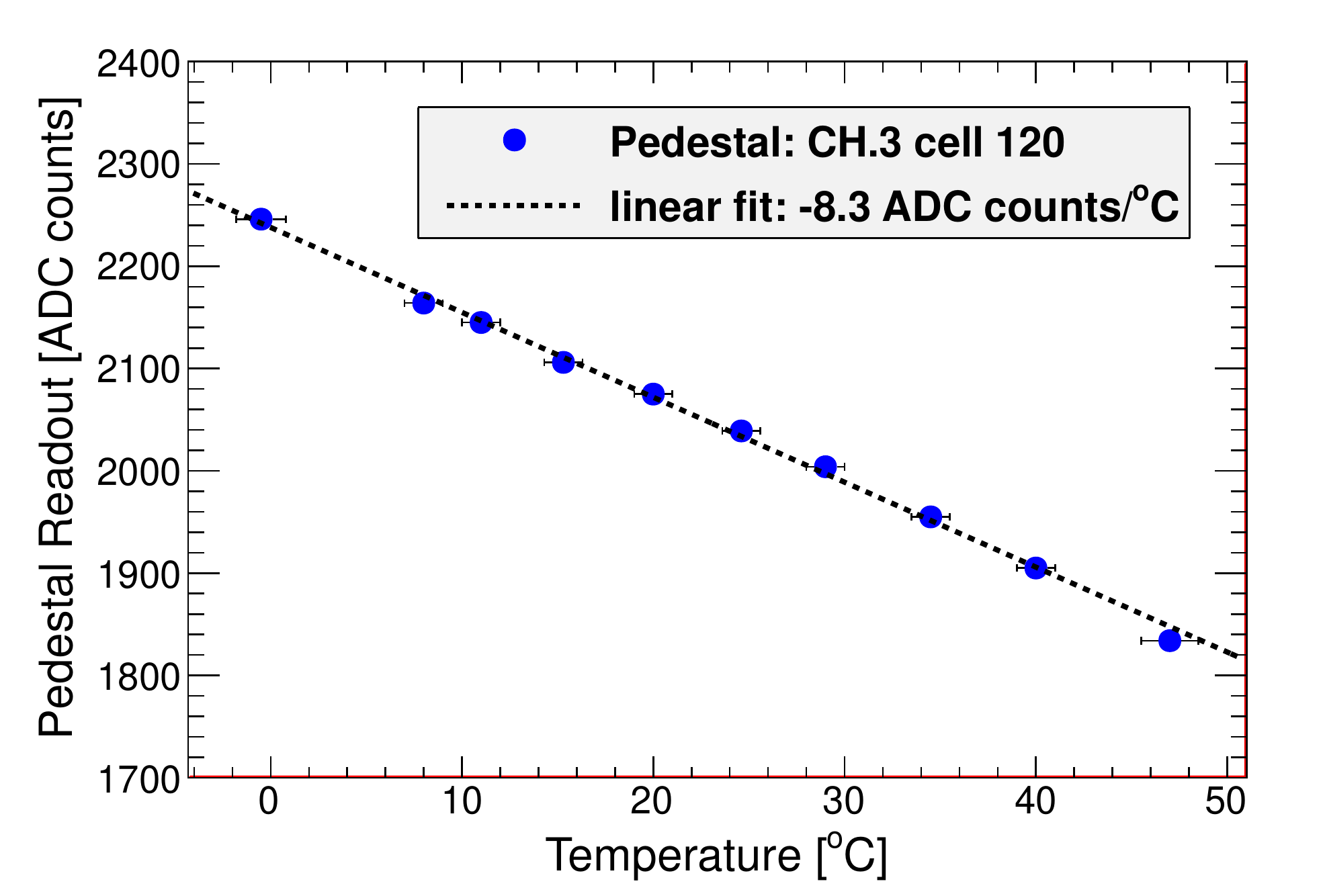}}
\caption{Temperature dependance of the pedestal level of a single cell. The data are consistent with a linear change in pedestal level over temperature.  }
\label{ped_temp}
\end{figure}

\begin{figure}[t!]
\centering
\subfloat[]{\includegraphics[scale=.44]{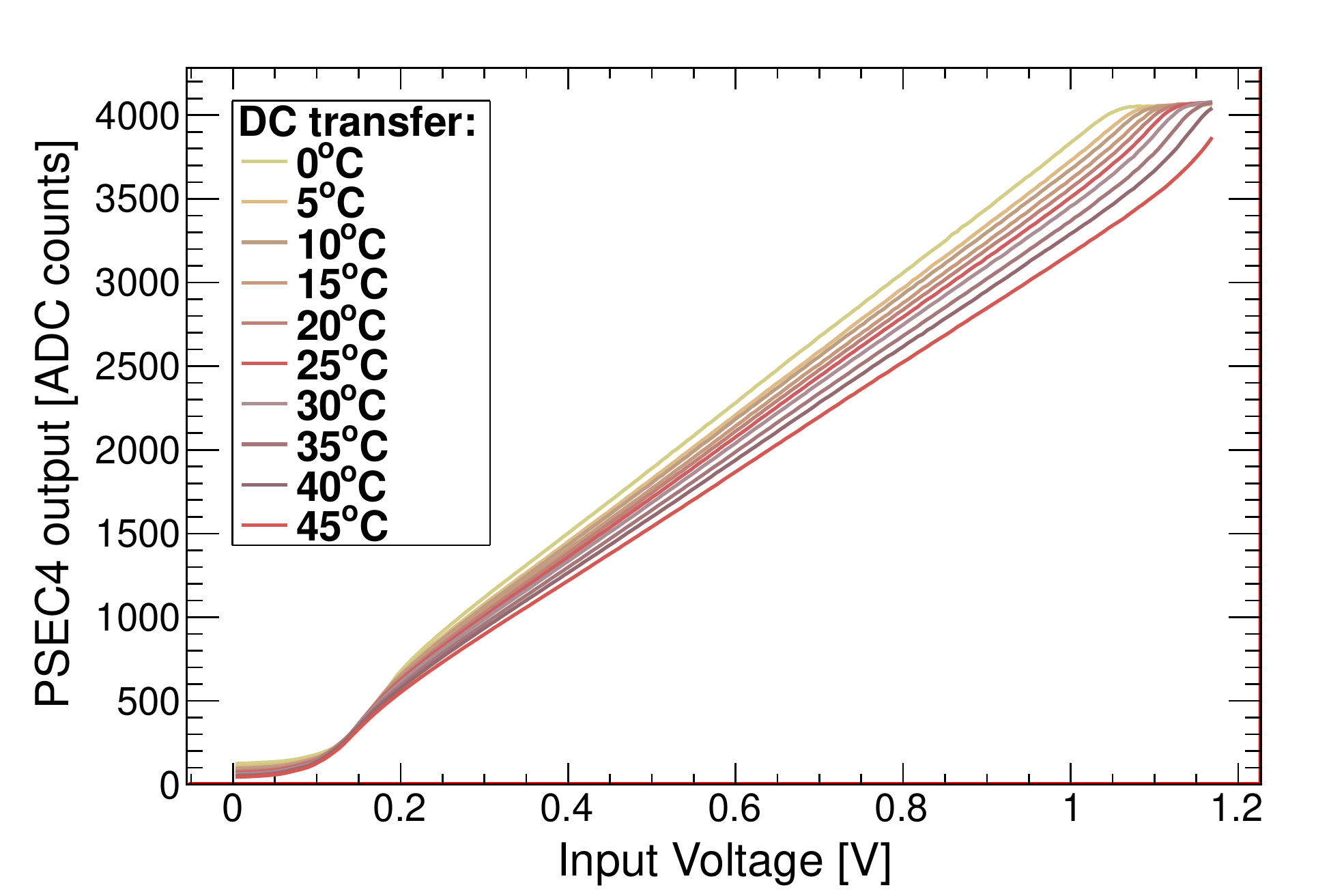}}%\qquad
\subfloat[]{\includegraphics[scale=.44]{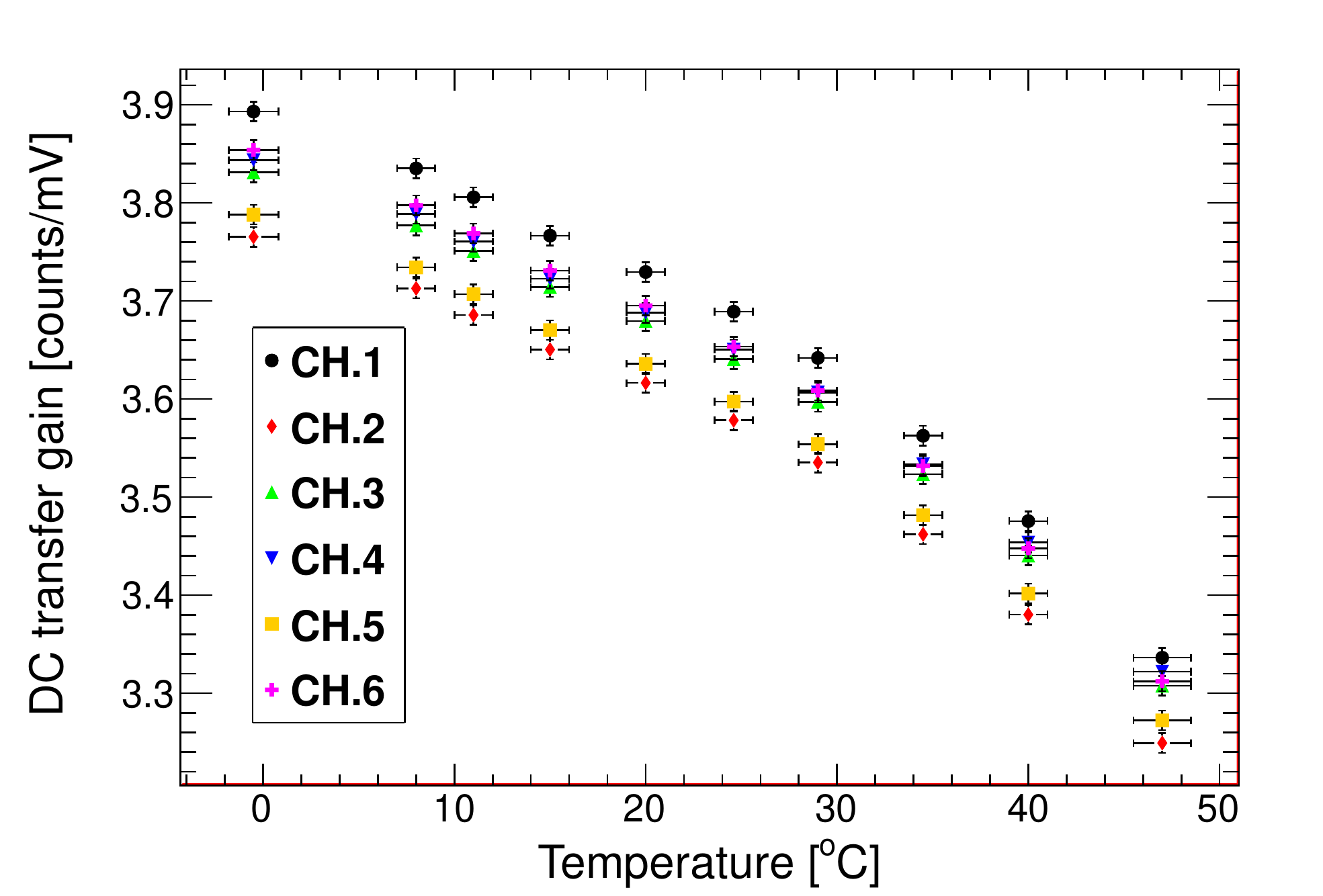}}
%\hspace{-20 pt}
\caption{(a)~Average DC response of a single PSEC4 channel over a temperature range of  0$^\circ$C to 45$^\circ$C. 
(b)~The extracted DC transfer gains as a function of temperature for all channels.}
\label{lin_temp}
\end{figure}

\subsubsection{Temperature Dependance}
%The temperature stability  of the PSEC4 ADC was measured.  
The ring oscillator ADC clock is the most temperature sensitive circuit and is servo-controlled using the FPGA to better than 0.1\% over a wide temperature range as shown in Figure~\ref{ro_temp}. 
The ADC clock frequency was measured using 50k events at each temperature.
Other temperature sensitive circuitry, including the chip-global ramp generator, are not feedback controlled. 

The mid-range cell pedestal temperature dependence is shown in Figure~\ref{ped_temp}. 
Pedestal levels are computed for each cell by recording the average ADC baseline over several readouts.
  The pedestal variation is consistent with a linear trend of 
$\sim$8.5~ADC counts/$^{o}$C. This trend is common to all cells in PSEC4.

The count-to-voltage transfer also shows temperature variation due to changes in the ADC ramp slope. 
The average DC transfer curves at different temperatures are shown in Figure~\ref{lin_temp}a. 
The count-per-voltage gain is extracted from a fit to the linear region of the DC transfer curve and is
plotted in Figure~\ref{lin_temp}b.
Since the ADC ramp is common to all channels, the average DC tranfer gains of all channels are observed to have
the same temperature dependence. To mitigate this effect, a feedback loop that serves the ramp current source could be
implemented.

\subsection{Sample Leakage}
\label{leakage_section}
When triggered, the write switch on each cell is opened and the sampled voltage is held 
at high impedance on the 20~fF capacitor (Fig.~\ref{psec_cell}). Two charge leakage pathways
are present: 1) sub-threshold conduction through the write switch 
formed by transistors T1 and T2; and 2) gate-oxide tunneling through the NFET at the comparator input.
The observable leakage current is the sum of these two effects.

\begin{figure}[t!]
\hspace{-30 pt}
\centering
\includegraphics[scale=.47]{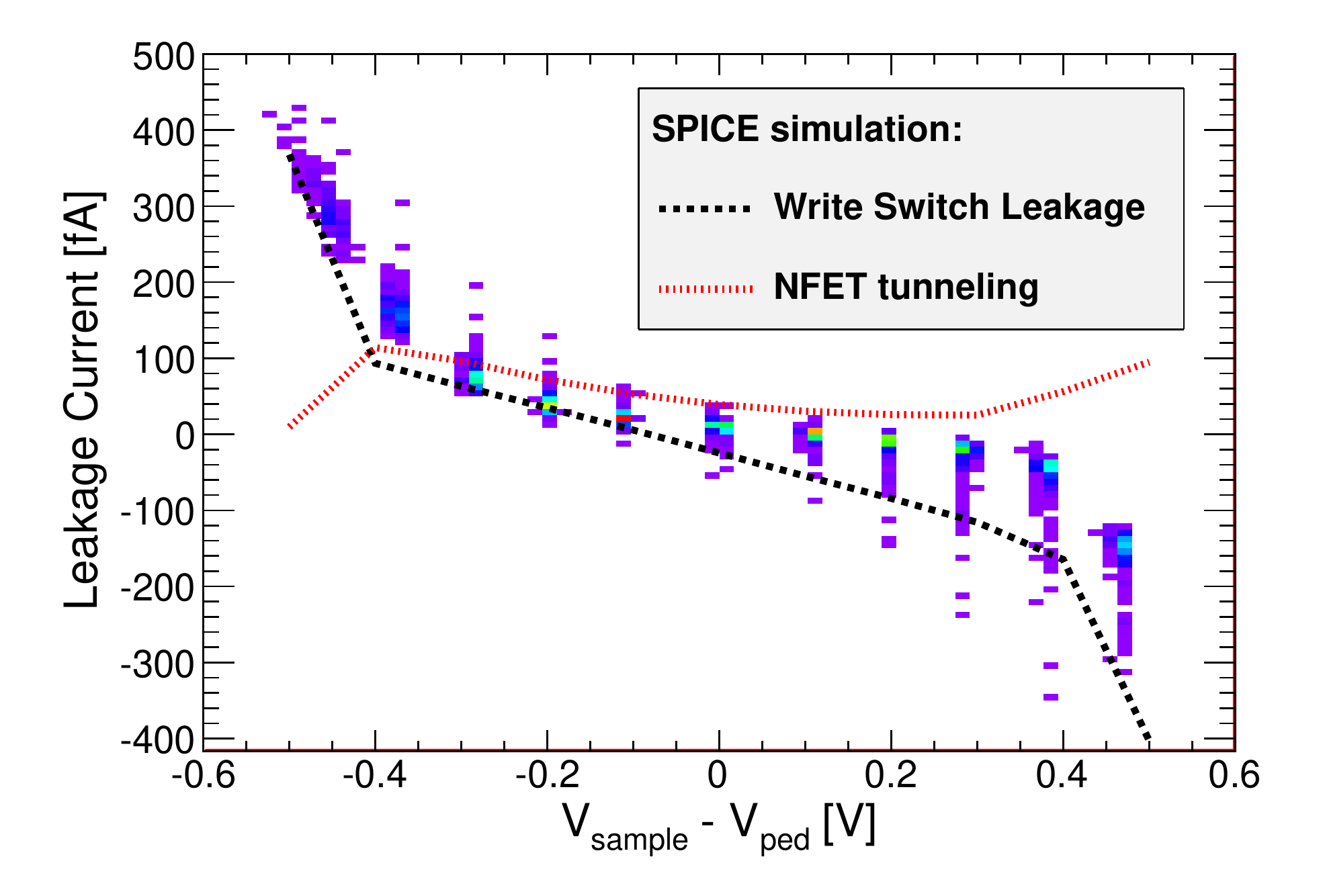}
\vspace{-10 pt}
\caption[sample leakage]{PSEC4 sample cell leakage measured at room temperature.
The measured leakage is shown by the histogrammed data points. Results from a 0.13~$\mu$m CMOS \textsc{spice} simulation
are also included. The simulation shows the leakage current contributions from 1) sub-threshold conduction through the disengaged write switch;
 and 
2)  gate-oxide tunneling from the NFET in the input stage of the comparator. }
\label{plot_leakage}
\end{figure}

To measure the leakage current, a 300~ns wide, variable-level pulse was sent to a single PSEC4 channel. 
Since the sampling window is 25~ns, each SCA cell sampled the transient level. 
After triggering, the sampled transient voltage was repeatedly digitized at 1~ms intervals
and the change in voltage on the capacitor was recorded over  a 10~ms storage-time.

The room temperature PSEC4 leakage current as a function of input voltage over the full 1~V dynamic range is shown in Figure~\ref{plot_leakage}. 
A pedestal level of V$_{DD}$/2~=~0.6~V was set at the input.  The measured leakage is shown in the 2-D histogram. A large 
spread (RMS~$\sim$70~fA) is seen at each voltage level. Results from a 0.13~$\mu$m CMOS \textsc{spice} simulation 
show that the write-switch leakage is the dominant pathway. A small amount ($\leq$100~fA)
of NFET gate-oxide tunneling is also consistent with the data. 
%The total leakage from these pathways qualitatively matches the data.

In normal operation, the ADC is started immediately after a trigger is registered. In this case, the analog voltage
hold time is limited to the ADC conversion time. Assuming a constant current, the leakage-induced voltage change is given by
\begin{equation}
\Delta V  = \frac{I_{leakage}\: \Delta t}{C_{sample}} 
\end{equation}
where $\Delta t$ is the ADC conversion time.
With the maximum leakage current of $\pm$500~fA and a conversion time of 4~$\mu$s, $\Delta V$ is $\pm$100~$\mu$V.
This value is at least 5$\times$ lower than the electronics noise.

\subsection{Noise}
\label{noise}
\begin{figure}[t!]
%\hspace{-20 pt}
%\includegraphics[scale=.44]{noise_4.pdf}
%\captionof{a}
%\includegraphics[scale=.44]{noise_temp.pdf}
%\captionof{b}
%\vspace{ -10 pt}
\centering
\subfloat[]{\includegraphics[scale=.4]{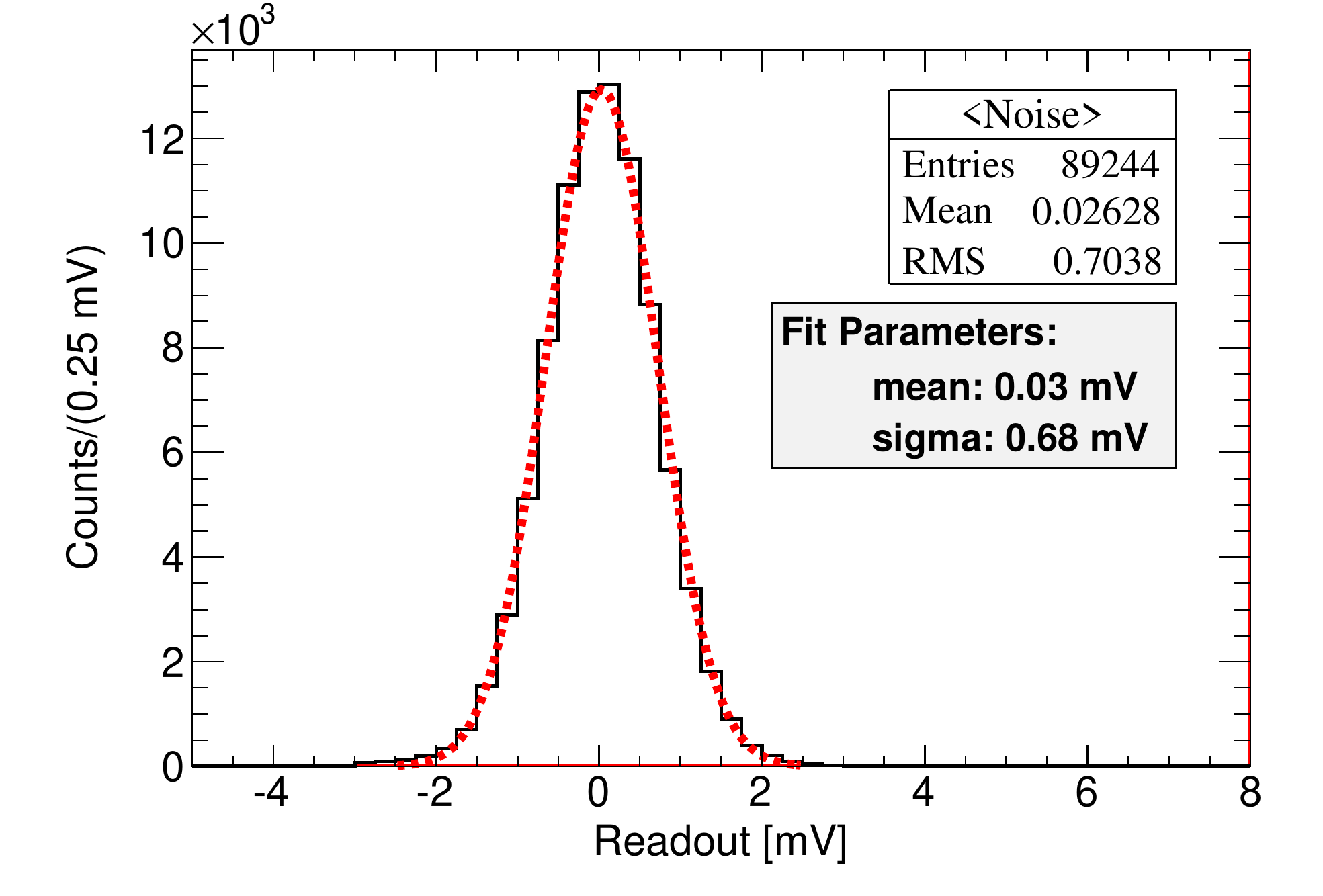}}%\qquad
\subfloat[]{\includegraphics[scale=.4]{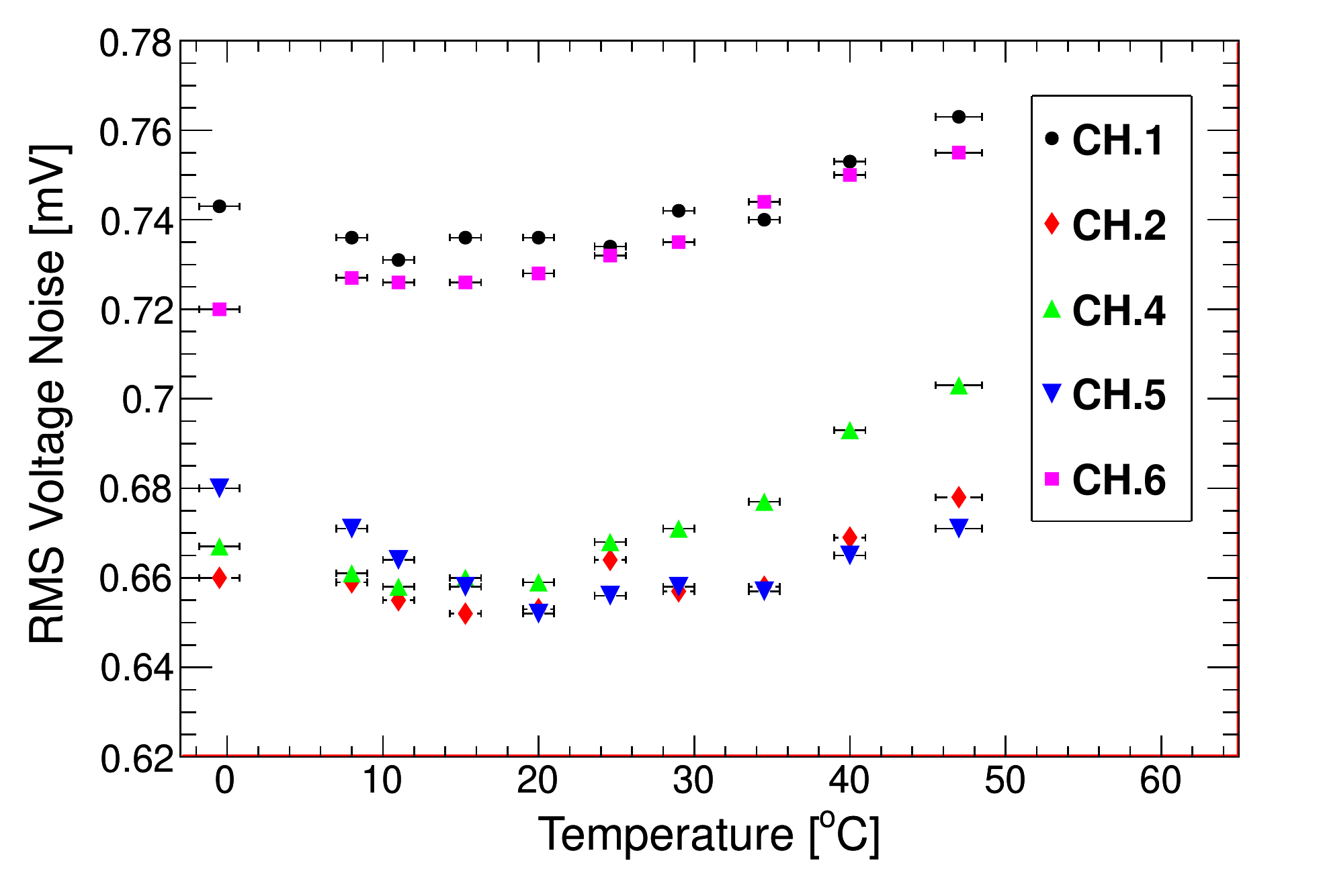}}
\caption[noise]{(a)~A PSEC4 baseline readout showing the electronic noise. The data are recorded 
from single channel after offset correction. 
The RMS value of $\sim$700~$\mu$V is representative of the room temperature electronics noise on all channels.
(b)~Channel RMS noise measured over temperature.  }
\label{noiseplot}
\end{figure}

After fixed-pattern pedestal correction and event-by-event baseline subtraction, which removes low-frequency noise contributions, 
the PSEC4 electronic noise is measured to be $\sim$700~$\mu$V RMS on all channels as shown in Figure~\ref{noiseplot}a. 
The noise level is consistently sub-mV over a $\pm$20$^\circ$C  temperature range around room temperature.
Above $\sim$20$^\circ$C, the electronics noise increases with temperature, which is 
consistent with the thermal noise expectation of~$\sqrt{k_BT/C}$ (Figure~\ref{noiseplot}b).
The noise figure is dominated by broadband thermal noise on the 20~fF sampling capacitor, 
%($\sqrt{k_BT/C_{sample}}$ )
which contributes 450~$\mu$V (RMS~60 electrons) at 300~K. 
Other noise sources include the ADC ramp generator and comparator. 
The noise corresponds to roughly 3 least significant bits (LSBs), reducing the DC RMS dynamic range
to 10.5~bits over the signal voltage range.

\begin{figure}[t!]
\vspace{-15 pt}
%\hspace{-15 pt}
\centering
\includegraphics[trim=85pt 0pt 0pt 0pt, clip=true,scale=0.33]{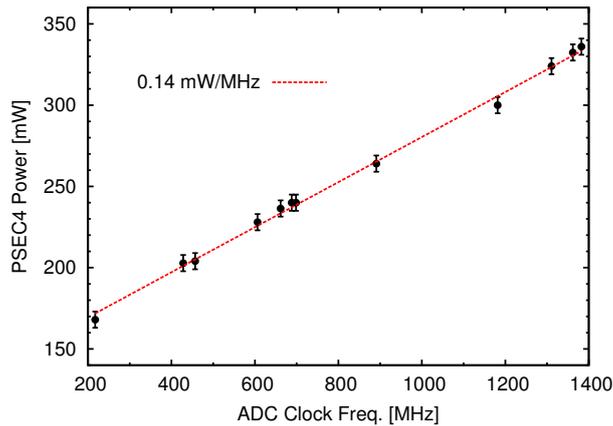}
\vspace{-22 pt}
\caption[Power]{ Total PSEC4 power as a function of the ADC clock rate. Clock rates between 200~MHz 
and 1.4~GHz can be selected based on the power budget and targeted ADC speed and resolution. When the ADC is not running, 
the quiescent (continuous sampling) power consumption is $\sim$40~mW per chip.}
\label{adcpower}
\end{figure}

\subsection{Power}
\label{power}
The power consumption is dominated by the ADC, which simultaneously clocks 1536 ripple counters
and several hundred large digital buffers at up to 1.4~GHz. The total power draw per chip as a 
function of ADC clock rate is shown in Figure~\ref{adcpower}.  
To reduce the steady state power consumption and to separate the chip's digital processes from 
the analog sampling, the ADC is run only after a trigger is sent to the chip.
Without a trigger, the quiescent power consumption is $\sim$40~mW per chip, including the
locked VCDL sampling at 10.24~GSa/s and the current biases of all the comparators.  

%Initiating the ADC with a clock rate of 1~GHz causes the power draw to increase from 40~mW 
%to 300~mW within a few nanoseconds.
%To mitigate high-frequency power supply fluctuations when switching on the ADC, several `large' (2 pF) 
%decoupling capacitors were placed on-chip near the ADC. 
%These capacitors, in addition with the close-proximity evaluation board decoupling capacitors ($\sim$0.1-10~$\mu$F), 
%prevent power supply transients from impairing chip performance.

At the maximum PSEC4 sustained trigger rate of 50~kHz, in which the ADC is running 20$\%$ of the time, a maximum
average power of 100~mW is drawn per chip.

\subsection{Frequency Response}
\label{freq}

\begin{figure}[t!]
\centering
\subfloat[]{\includegraphics[scale=.3]{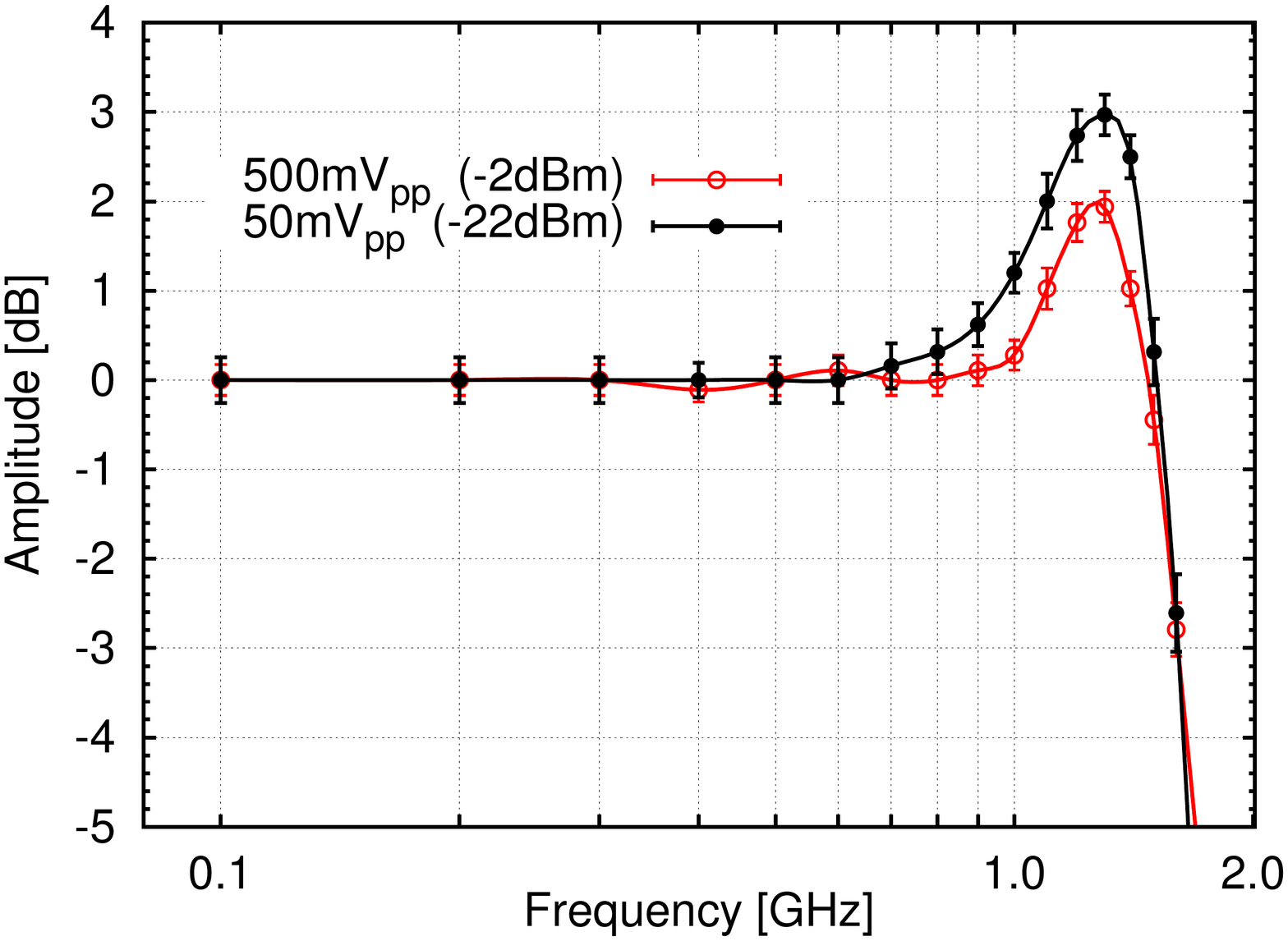}}
\subfloat[]{\includegraphics[scale=.3]{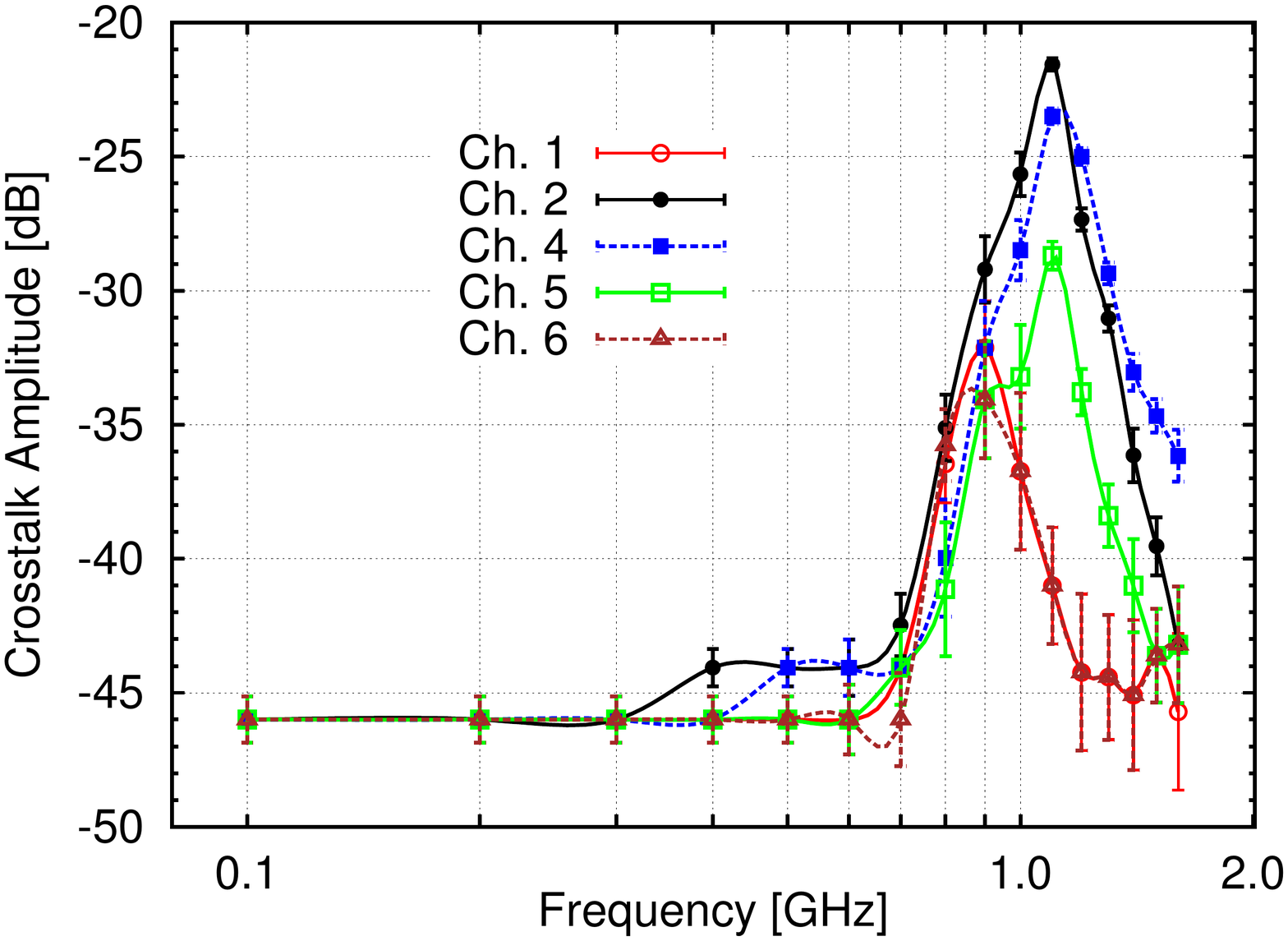}}
%\vspace{ -20 pt}
\caption[ABW]{(a)~PSEC4 frequency response. The -3~dB analog bandwidth is 1.5~GHz. The positive resonance above 1~GHz is due to 
bondwire inductance of the signal wires in the chip package. Similar responses are shown for 
large and small sinusoidal inputs.
(b)~Channel-to-channel crosstalk as a function of frequency. Channel~3 was driven with a -2~dBm sinusoidal input. Adjacent channels see a maximum of -20~dB crosstalk at 1.1~GHz. The electronic noise floor is -50~dB for reference. }
\label{analogbw}
\end{figure}

 The target analog bandwidth for the PSEC4 design was $\geq$1 GHz. 
The bandwidth is limited by the parasitic input capacitance (\textit{C$_{in}$}), which
drops the input impedance at high frequencies\footnote{This ignores 
negligible contributions to the impedance due to the sampling cell input coupling. 
The write switch on-resistance ($\leq$ 4~k$\Omega$ over the full dynamic range) 
and the 20~fF sampling capacitance introduce a pole at $\geq$2~GHz.} as
\begin{equation}
| Z_{in} | = \frac{R_{term}}{\sqrt{1 + \omega^2\:R_{term} \: C_{in}}}
\end{equation}
where \textit{R$_{term}$} is an external 50~$\Omega$ termination resistor. 
Accordingly, the expected half-power bandwidth is given by:
\begin{equation}
f_{3dB} = \frac{1}{2 \pi  \:  R_{term} \: C_{in}}
\end{equation}

%\begin{figure}[t]
%\hspace{-40 pt}
%\centering
%\includegraphics[scale=.33]{plotxt.pdf}
%\vspace{ -20 pt}
%\caption[xtalk]{The channel-to-channel crosstalk as a function of frequency. Channel~3 was driven with a -2~dBm sinusoidal input. Adjacent channels see a maximum of -20~dB crosstalk at 1.1~GHz. The electronic noise floor is -50~dB for reference. }
%\label{xtalkfig}
%\end{figure}

The extracted \textit{C$_{in}$} from post-layout studies was $\sim$2~pF,
 projecting a -3~dB bandwidth of 1.5~GHz which corresponds to the measured value shown in
Figure~\ref{analogbw}a. The chip package-to-die bondwire inductance gives
a resonance in the response above 1~GHz that distorts 
signal content at these frequencies. An external filter may be added to flatten
the response.

The measured channel-to-channel crosstalk is -25~dB below 1~GHz for all channels as shown 
in Figure~\ref{analogbw}b. 
For frequencies less then 700~MHz, this drops to better than -40~dB.
The primary crosstalk mechanism is thought to be the mutual inductance between signal bondwires in the chip package.
High frequency substrate coupling on the chip or crosstalk between input traces on the PSEC4 evaluation board may also contribute.

%\subsubsection{AC linearity}
%The PSEC4 response to sinusoidal signals of varying magnitude is shown in Figure~\ref{aclinear}. 
%The AC signals were calibrated using the DC transfer count-to-voltage LUT (Figure~\ref{adclinear}).
%Saturation is observed for signals as low as 100~MHz ($\sim$9\% above V$_{pp}$ of 500~mV) and this effect becomes constant for frequencies above 500~MHz ($\sim$15\% above V$_{pp}$ of 500~mV).

%This AC saturation effectively reduces the PSEC4 ADC dynamic range. With a DC calibration, it was shown that 10.5 bits effectively covered a 1~V range including electronics noise. At 100 MHz, the dynamic range drops to $\sim$10.35 bits. For signals up to 1~GHz, the dynamic range is 10.3 bits. 
%Additionally, an AC count-to-voltage conversion calibration LUT can be used for correct for this saturation.

%\begin{figure}[t!]
%\hspace{-46 pt}
%\centering
%\includegraphics[scale=.45]{aclin2.pdf}
%\vspace{-5 pt}
%\caption[ADC linearity]{AC response of a single channel. The PSEC4 data were converted to voltage using the DC linearity LUT as shown in Fig.~\ref{adclinear}. Saturation of the AC signals is observed for all frequencies, most clearly above 500~mV$_{pp}$}
%The AC saturation effect becomes constant at frequencies above 500~MHz.}
%\label{aclinear}
%\end{figure}

%%%%%%%%%%%%%%%%%%%%%%%%%%%%%
\subsection{Sampling Calibration}
\label{timebase}

For precision waveform feature extraction, both the overall time-base of the VCDL and
the cell-to-cell time step variations must be calibrated. With the rate-locking DLL, 
the overall PSEC4 sampling time base is stably servo-controlled at a default rate of 10.24~GSa/s.
The time-base calibration of the individual 256~delay stages, which vary due to cell-to-cell transistor size
mismatches in the VCDL, is the next task. Since this is a fixed-pattern variation, the time-base calibration
is typically a one-time measurement.

The brute force `zero-crossing' time-base calibration method is employed~\cite{kurtis}. 
This technique counts the number of times a sine wave input crosses zero voltage at each sample cell.
With enough statistics, the corrected time per cell is extracted from the number of zero-crossings (\textit{N$_{zeros}$})
using

\begin{equation}
<\Delta t> = \frac {T_{input}<N_{zeros}>}{2 \:N_{events}}
\end{equation}
where \textit{T$_{input}$} is the period of the input and \textit{N$_{events}$} is the number of
digitized sine waveforms. A typical PSEC4 time-base calibration uses 10$^6$~recorded events of 400 MHz sinusoids.

\begin{figure}[HtH!]
\centering
\subfloat[]{\includegraphics[scale=.44]{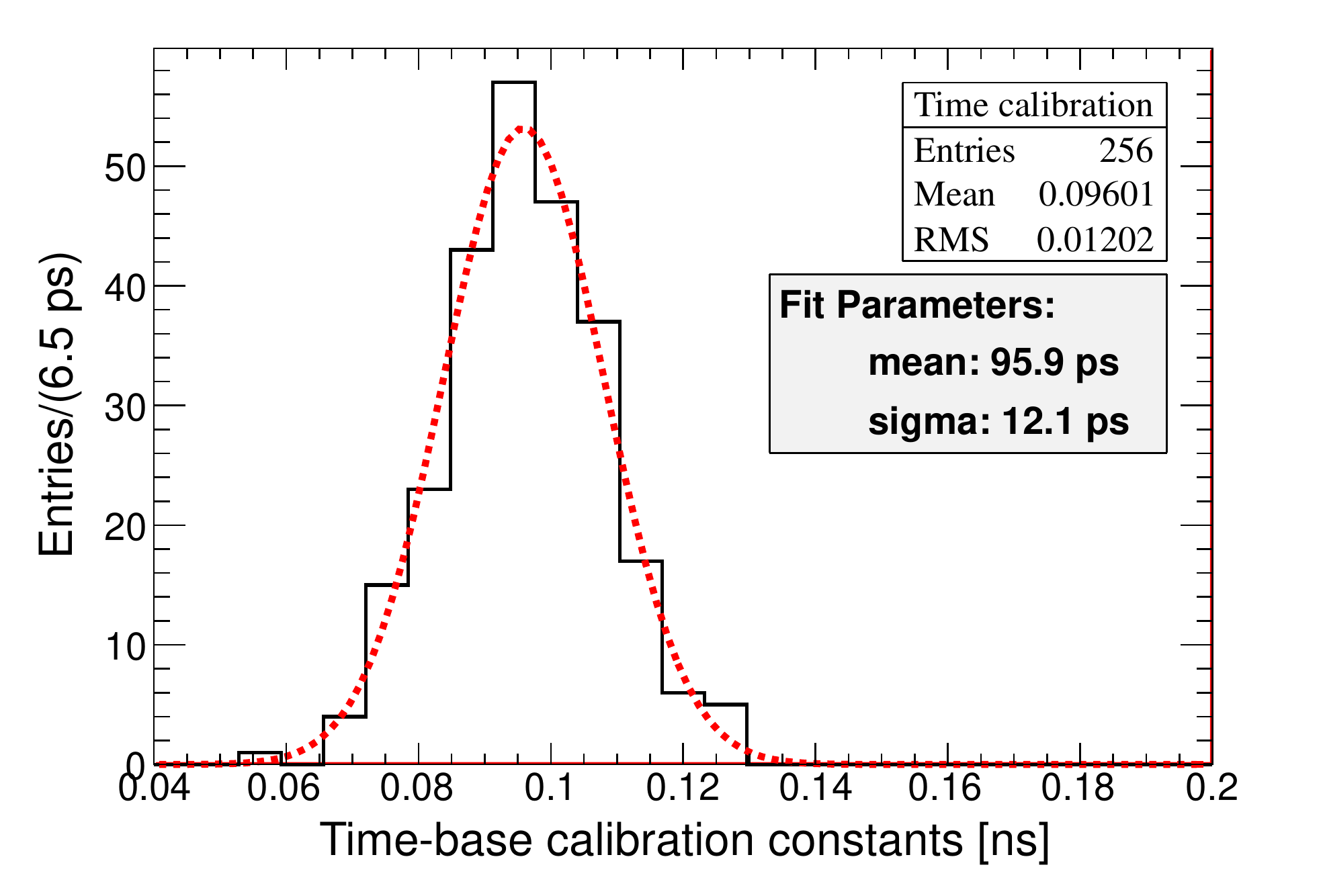}}
\subfloat[]{\includegraphics[scale=.44]{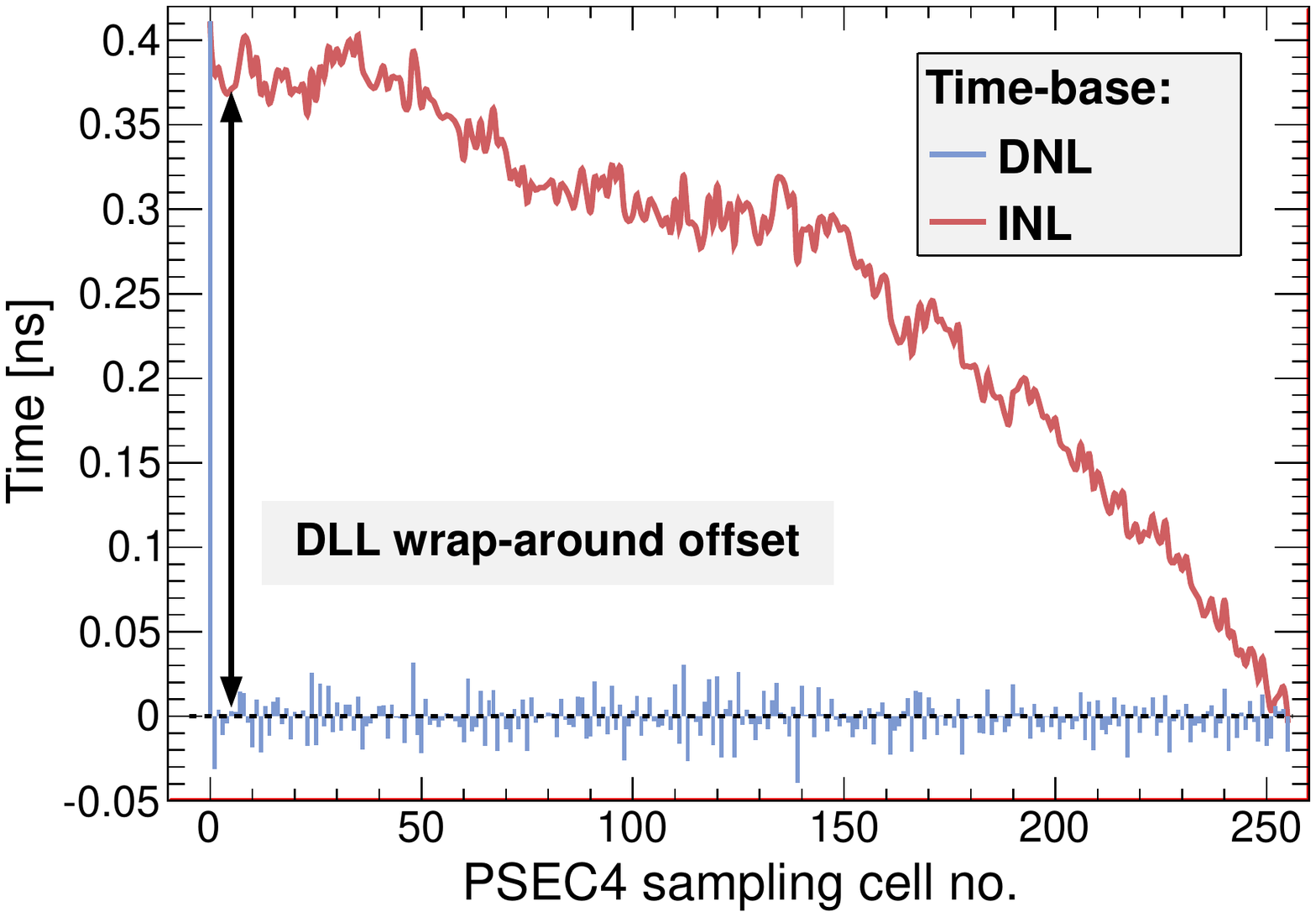}}

\caption[timebase calibration]{(a)~A histogram of the extracted time-base calibration constants ($\Delta$t) from a single channel.
These values are calculated using the zero-crossing technique and are used to correct the sampling time-base of the PSEC4 chip.
%A 13$\%$ spread in the $\Delta$t values is observed. 
%The average sampling rate over these cells is found to be $\sim$10.4~GSa/s, slightly higher than the nominal value.
(b)~The differential (DNL) and integral non-linearity (INL) of the PSEC4 time-base. The extracted $\Delta$t's are compared to an 
ideal linear time-base with equal time-steps per sample point. 
The large time-step at the first sample bin is caused by a fixed DLL latency when wrapping the sampling 
from the last cell to the first.} %With the servo-locking DLL the INL is constrained to be zero at the last cell.}
\label{time_cal}
\end{figure}

\begin{figure}[t!]
%\hspace{-20 pt}
\centering
\includegraphics[scale=.45]{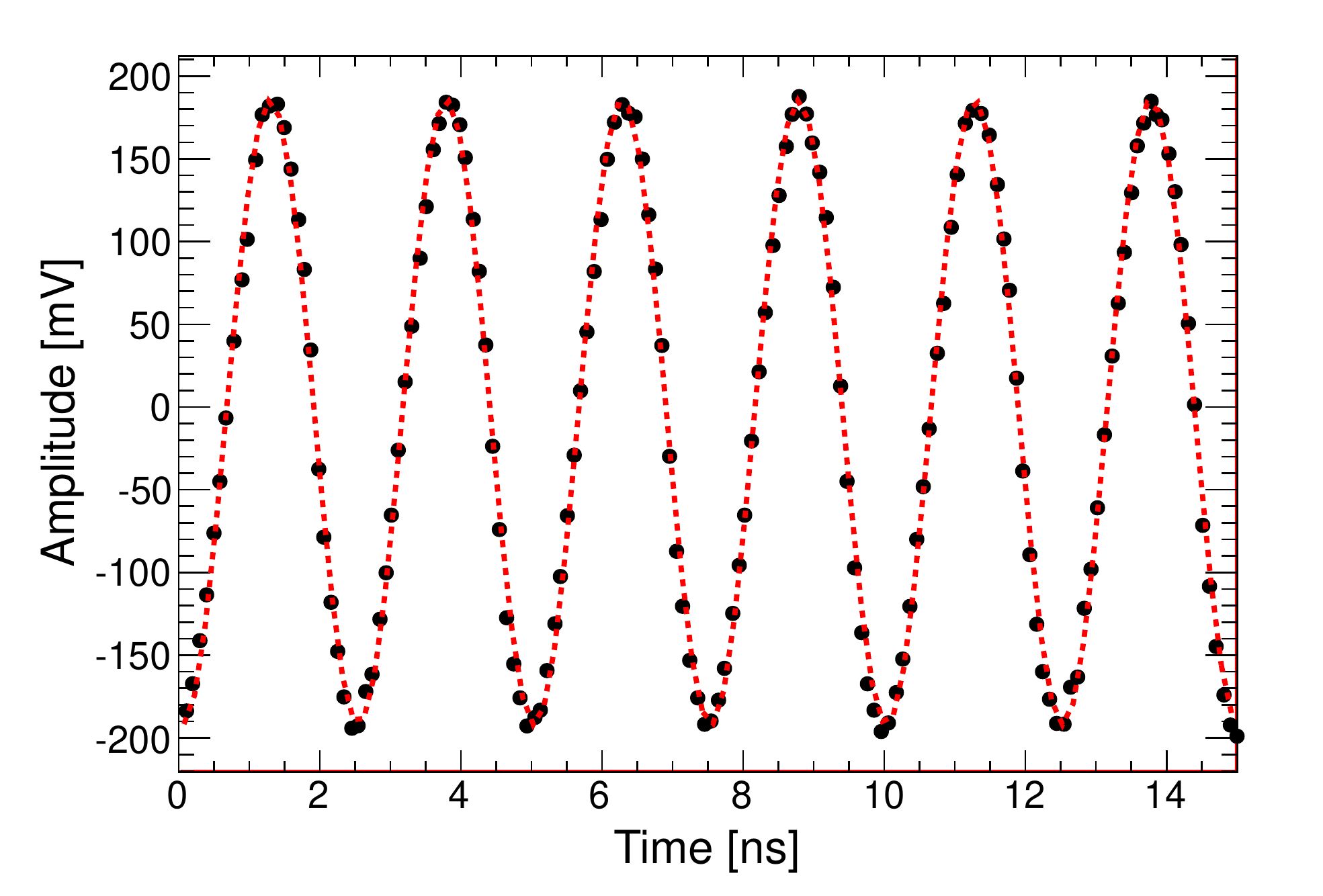}
\vspace{ -9 pt}
\caption[sine after calibration]{A 10.24~GSa/s capture of a 400~MHz sine input is shown (black dots) after a
channel-only linearity correction and time-base calibration. A fit (red dotted line) is applied to the data. }
\label{psec_sine}
\end{figure}

\begin{figure}[t!]
%\hspace{-20 pt}
\centering
\includegraphics[scale=.46]{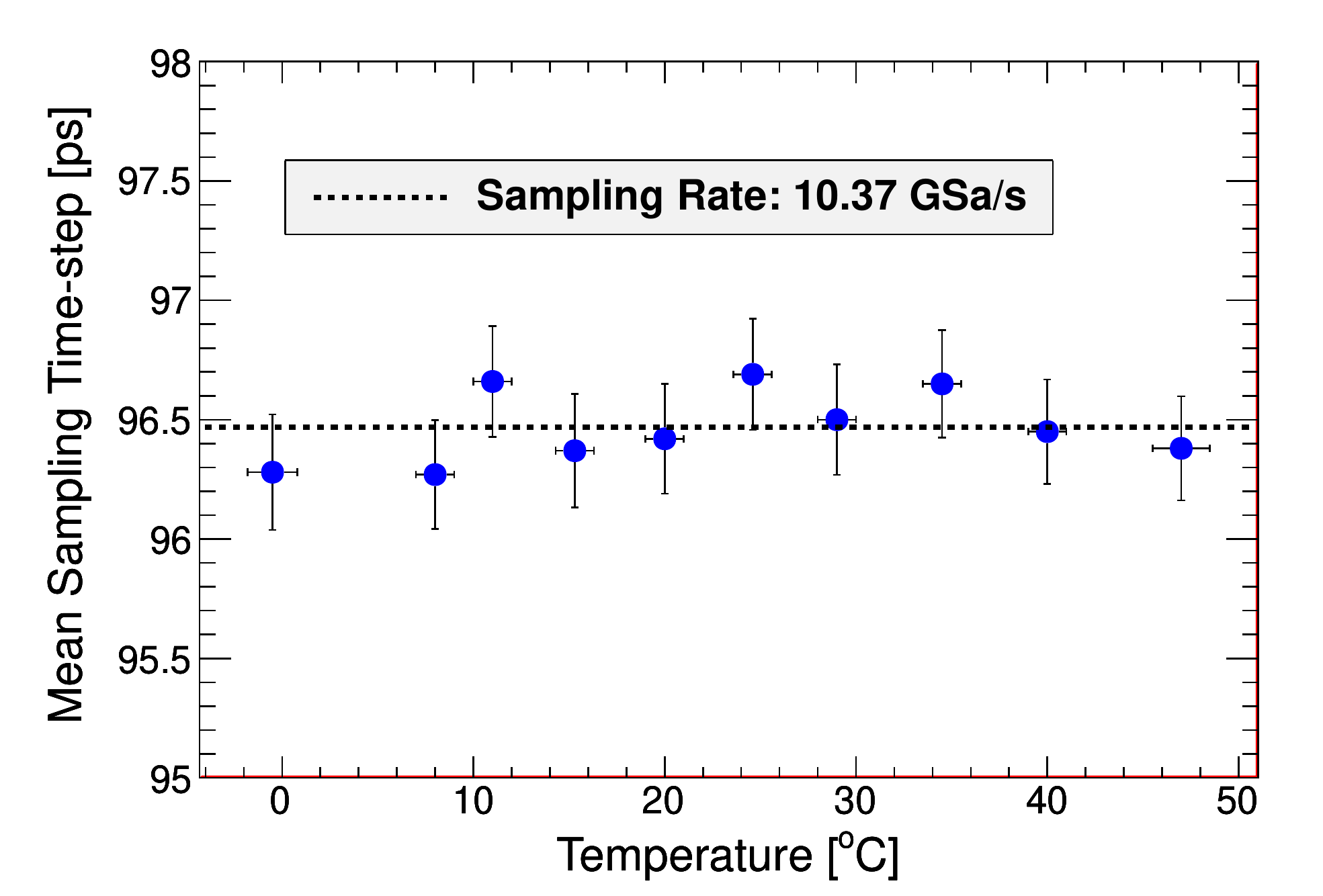}
\vspace{ -9 pt}
\caption[temp sample]{Mean time-base step as measured over a temperature range -1$^\circ$C to 46$^\circ$C. The sampling rate
is held constant over temperature with the on-chip DLL.}
\label{sample_temp}
\end{figure}

The variation of the time-base sampling steps is $\sim$13$\%$ as shown in Figure~\ref{time_cal}a. 
Due to a relatively large time step at the first cell, the average sampling rate over the remaining 
VCDL cells is $\sim$10.4~GSa/s, slightly higher than the nominal rate.  
With the servo-locking DLL the INL is constrained to be zero at the last cell.
A digitized 400~MHz sine wave is shown in Figure~\ref{psec_sine} after applying the time-base calibration constants.

The non-linearity of the PSEC4 time-base is shown in Figure~\ref{time_cal}b. 
Each bin in the plot is indicative of the 
time-base step between the binned cell 
and its preceding neighbor cell. 
The relatively large DNL in the first bin, which corresponds to the delay
between the last (cell~256) and first sample cells, is caused by a fixed DLL latency when wrapping the sampling from the last cell to the first.

The mean sampling rate is shown in Figure~\ref{sample_temp} to be uniform over temperature. 
This is an expected feature of the on-chip DLL.
Fifty-thousand events were
recorded at each temperature point and the zero crossing algorithm was run on each dataset. No temperature
dependence is observed.

\subsection{Waveform Timing}
\label{timing}
The effective timing resolution of a single measurement is calculated by waveform feature extraction after
linearity and time-base calibration. A 0.5~V$_{pp}$, 1.25~ns FWHM Gaussian pulse was created
using a 10~GSa/s arbitrary waveform generator (Tektronix AWG5104). The output of the AWG 
was sent to 2~channels of the ASIC using a broadband-RF 50/50 splitter.
Once digitized, an off-line Gaussian functional fit is performed to the leading edge of the pulse.
The pulse times from both channels are extracted from the fit and are subtracted on an event-by-event
basis.

\begin{figure}[t!]
\centering
\subfloat[]{\includegraphics[scale=.42]{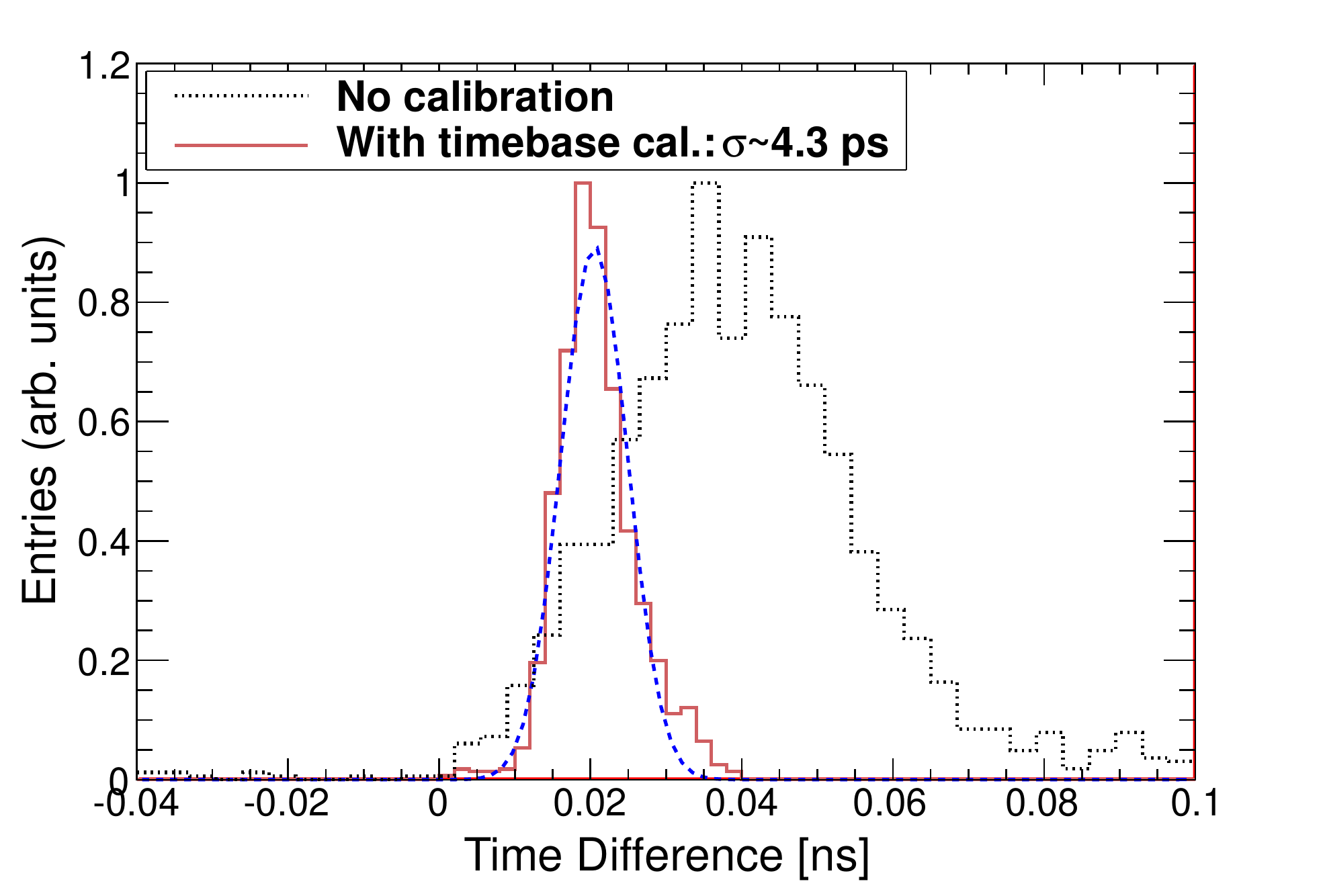}}
\subfloat[]{\includegraphics[scale=.42]{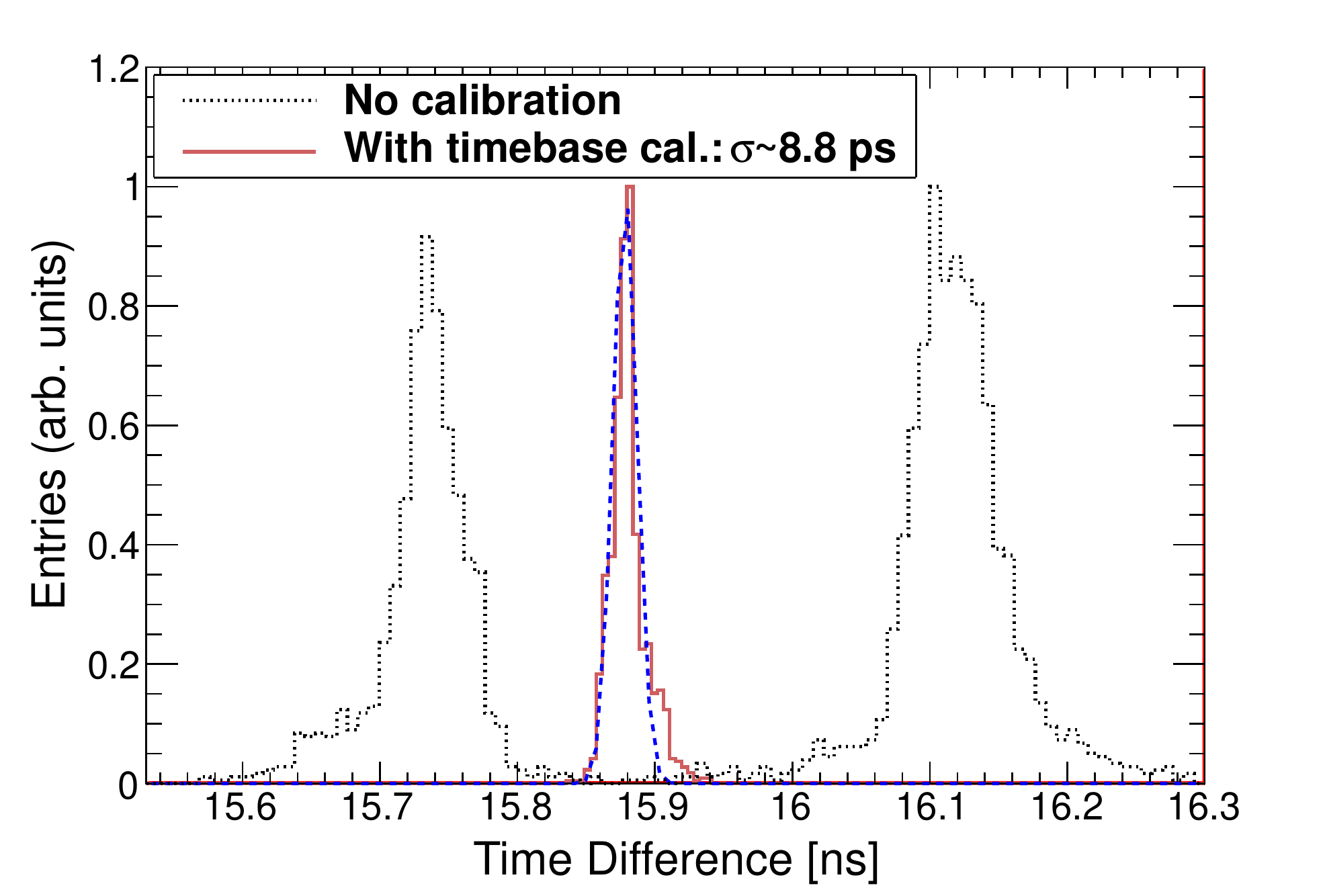}}
\caption{Time resolution results before (black-dashed) and after (maroon-solid lines) applying timebase calibrations. Pulses were injected asynchronously into 2 channels of PSEC4. Data are shown for pulse separations of (a)~$\sim$0~ns, and (b)~$\sim$16~ns. The bimodal distribution in the pre-calibration data of (b) corresponds to the fixed DLL wraparound offset. }
\label{time_hists}
\end{figure}

The timing resolution was measured by asynchronous pulse injection to two channels of PSEC4.  
In this measurement, the two pulses were delayed relative to one-another and the waveforms were
captured uniformly across the 25~ns PSEC4 sampling buffer. 
Ten-thousand events were recorded at each delay stage. 
The time difference resolution was extracted before and after applying the time-base calibration to both channels.
Figures~\ref{time_hists}a and~\ref{time_hists}b show the timing resolution results for pulse separations in the PSEC4 buffer of 0~ns and~16~ns, respectively. For uncalibrated data in which the pulse separation is non-zero, the bimodal timing distribution is due to the fixed DLL-wraparound offset. 

\begin{figure}[t!]
\centering
\includegraphics[scale=.42]{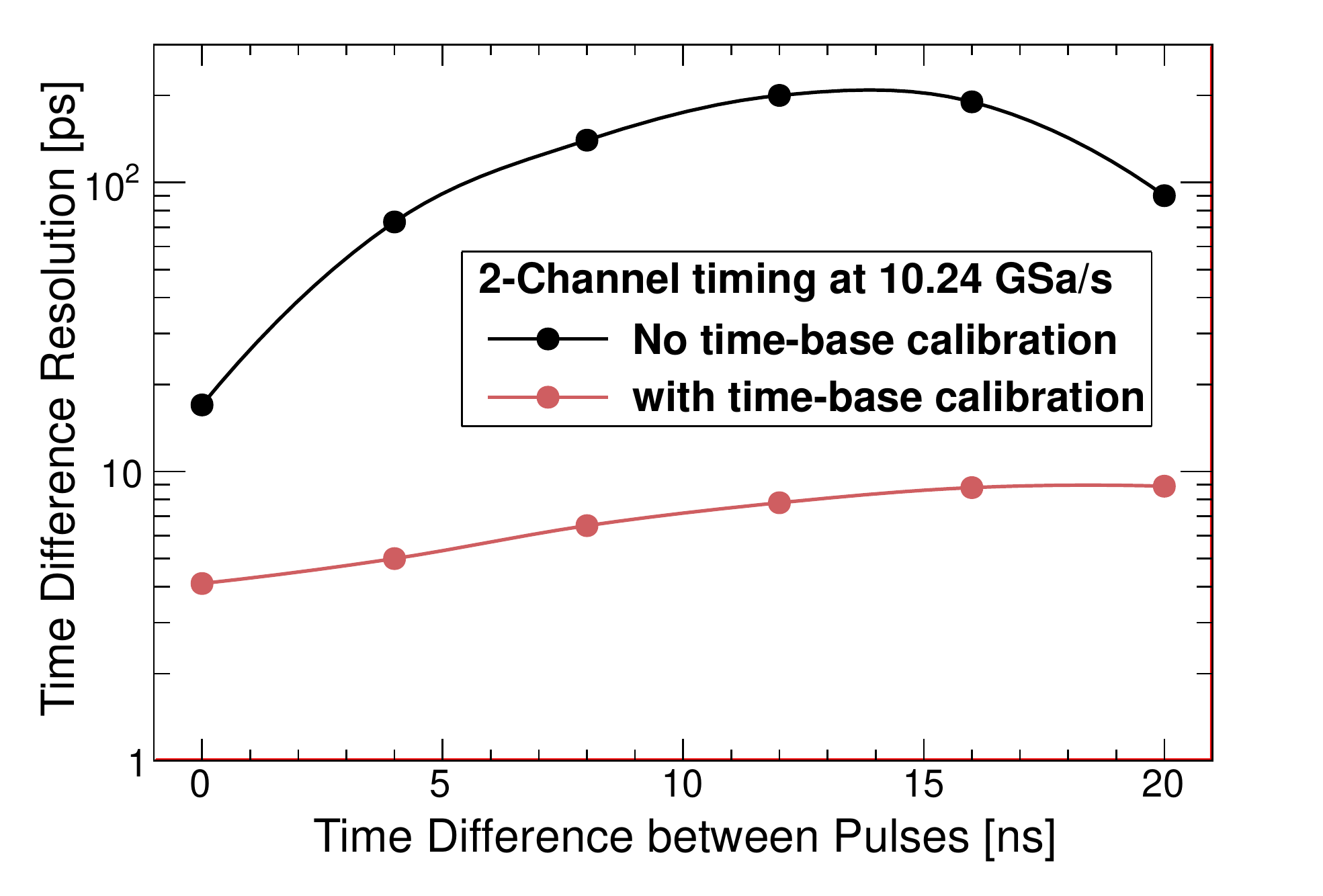}
\caption{Two-channel time resolution from asychronous pulse injection into PSEC4. 
The timing resolution as a function of pulse separation in the PSEC4 sampling buffer is shown.}
\label{time_res}
\end{figure}

A PSEC4-calibrated timing resolution of 9~ps or better was measured over a 20~ns span of pulse separation as shown in Figure~\ref{time_res}. The 1$\sigma$ resolution was extracted by fitting the calibrated timing distributions shown in Figure~\ref{time_hists}.  For simultaneous pulses the resolution is $\sim$4~ps, but the timing resolution slightly degrades as the time difference between pulses is increased. The large RMS in the timing for uncalibrated data is due to the DLL wraparound offset of roughly 400~ps.

\subsection{Performance Summary}
\label{perf_table}
The performance and key architecture parameters of PSEC4 are summarized in Table~\ref{table}.

%%%%%%%%%%%%%%%%%%%%%%%%%%%%
\section{Conclusions}
\label{conclusion}
We have described the architecture and performance of the PSEC4 waveform digitizing ASIC.
The advantages of implementing waveform sampling IC design in a deeper sub-micron process
are shown, with measured sampling rates of up to 15 GSa/s and analog bandwidths of 1.5 GHz. Potential
0.13~$\mu$m design issues, such as leakage and dynamic range, were optimized and provide a 
1~V dynamic range with sub-mV electronics noise. After a one-time timebase calibration,
it is possible to extract precision timing measurements (4 - 9 psec) when applying a simple rising-edge fit to the PSEC4 
digitized waveform.
%The first application of the PSEC4 ASIC is the compact,low-power front-end waveform sampling of LAPPD MCP-PMTs.

\begin{table*}[]
\caption{PSEC4 architecture parameters and measured performance results.}
\vspace{5 pt}
\label{table}
  \centering
  \begin{tabular}{l|l|l}
    \hline
    Parameter &   Value  &  Comment \\
    \hline
    \hline
    Channels & 6 & die size constraint \\ 
    Sampling Rate & 4-15 GSa/s & servo-locked on-chip\\ 
    Samples/channel & 256 & 249 samples effective\\
    Recording Buffer Time & 25 ns & at 10.24~GSa/s \\
    Analog Bandwidth & 1.5 GHz &  \\
    Crosstalk &  7\%  &  max. over bandwidth\\
     & $\textless$1\%  & typical for signals $\textless$800~MHz \\
    Noise & 700 $\mu$V & RMS (typical). RF-shielded enclosure. After calibration. \\
    DC RMS Dynamic Range & 10.5 bits & 12 bits logged. Linearity calibration for each cell.\\
    Signal Voltage Range & 1~V & after linearity correction \\
    ADC conversion time & 4 $\mu$s &  max. 12 bits logged at 1~GHz clock speed \\
               & 250 ns  & min. 8-bits logged at 1~GHz \\
    ADC clock speed & 1.4~GHz & max. \\
    Readout time & 0.8$\textit{n}$~$\mu$s & $\textit{n}$ is number of 64-cell blocks to read ($\textit{n}$ = 24 for entire chip)\\
    Sustained Trigger Rate & 50~kHz & max. per chip. Limited by [ADC time + Readout time]$^{-1}$\\
    Power Consumption & 100~mW &  max. average power\\
    Core Voltage & 1.2~V & 0.13~$\mu$m CMOS standard \\
    \hline
    \hline
  \end{tabular}
\end{table*}

\section{Acknowledgements}
\label{acknowledgements}
We thank Mircea Bogdan, Fukun Tang, Mark Zaskowski, and Mary Heintz for
their strong support in the Electronics Development Group of the
Enrico Fermi Institute.  Stefan Ritt,
Eric Delagnes, and Dominique Breton provided invaluable guidance and
advice on SCA chips. We thank Kostas Kloukinas and the MOSIS educational program
for support and advice on the 0.13~$\mu$m CMOS process and project submissions

This work is supported by the Department of Energy, Contract
Nos. DE-AC02-06CH11357 and DE-SC0008172 , and the National Science Foundation, Grant No. 
PHY-106 601 4.

\end{document}